\documentclass[journal]{IEEEtran}
\usepackage{cite}
\usepackage{amsmath,amssymb,amsfonts}
\usepackage{enumerate}
\usepackage{xcolor}
\usepackage{mathtools,nccmath}
\usepackage{tabularx} 
\usepackage{graphicx}

\newcommand{\avec}{{\bf{a}}}

\newcommand{\evec}{{\bf{e}}}

\newcommand{\yvec}{{\bf{y}}}

\newcommand{\wvec}{{\bf{w}}}
\newcommand{\xvec}{{\bf{x}}}

\newcommand{\onevec}{{\bf{1}}}
\newcommand{\zerovec}{{\bf{0}}}

\newcommand{\Lambdamat}{{\bf{\Lambda}}}

\newcommand{\Amat}{{\bf{A}}}

\newcommand{\Emat}{{\bf{E}}}

\newcommand{\Imat}{{\bf{I}}}
\newcommand{\Lmat}{{\bf{L}}}

\newcommand{\Omat}{{\bf{O}}}

\newcommand{\Smat}{{\bf{S}}}

\newcommand{\Umat}{{\bf{U}}}

\newcommand{\Wmat}{{\bf{W}}}

\newcommand{\define}{\stackrel{\triangle}{=}}

\def\bsigma{{\mbox{\boldmath $\Sigma$}}}

\newcommand{\be}{\begin{equation}}
\newcommand{\ee}{\end{equation}}
\newcommand{\beqna}{\begin{eqnarray}}
\newcommand{\eeqna}{\end{eqnarray}}

\usepackage{subfig}
\usepackage{commath}
\usepackage{amsthm}
\usepackage{xcolor}
\usepackage{algorithmic}
\usepackage[linesnumbered,ruled,vlined]{algorithm2e}
\SetAlgoSkip{}

\newtheorem{proposition}{Proposition}

\newtheorem{mydef}{Definition}
\usepackage{url}
\DeclareMathOperator*{\argmax}{arg\,max}

\SetCommentSty{mycommfont}

\SetKwInput{KwInput}{Input}                
\SetKwInput{KwOutput}{Output}              

\newcommand\numberthis{\addtocounter{equation}{1}\tag{\theequation}}
\begin{document}
\title{Identification of Edge Disconnections in Networks Based on Graph Filter Outputs
	\thanks{{\footnotesize{S. Shaked and T. Routtenberg are with the School of Electrical and Computer Engineering Ben-Gurion University of the Negev Beer-Sheva 84105, Israel, e-mail: shlomsh@post.bgu.ac.il, tirzar@bgu.ac.il.}}}
\thanks{This work is supported by the  Israeli Ministry of National Infrastructure, Energy and Water Resources and  by the BGU Cyber Security Research Center.\\
This work has been submitted to the IEEE for possible publication.
Copyright may be transferred without notice, after which this version may
no longer be accessible.
}
}
\author{Shlomit~Shaked and
Tirza~Routtenberg,~\IEEEmembership{Senior~Member,~IEEE,}
}

\maketitle

\begin{abstract}
Graphs are fundamental mathematical structures
used in various fields to model statistical and physical relationships between data, signals, and processes. In some applications, such as data processing in graphs that represent physical networks, the initial network topology is known. However, disconnections of edges in the network change the topology and may affect the signals and processes over the network. In this paper, we consider the problem of  edge disconnection identification in networks by using concepts from graph signal processing (GSP). We assume that  the graph signals measured over the vertices of the network can be represented as white noise that has been filtered on the graph topology by a smooth graph filter.
 We develop the likelihood ratio test (LRT) to detect a specific set of edge disconnections. 
 Then, we provide the  maximum likelihood (ML) decision rule for identifying general scenarios of edge disconnections in the network. It is shown that the sufficient statistics of the LRT and ML decision rule are the graph-frequency energy levels in the graph spectral domain. However, the ML decision rule
leads to a high-complexity exhaustive search over the edges in the network and is practically infeasible.
 Thus, we propose a low-complexity greedy method that identifies  a single disconnected edge at each iteration.
 Moreover, by using the smoothness  of the considered graph filter,  we  suggest a local implementation of the decision rule, which is
 based solely  on the measurements at neighboring vertices.
 Simulation results demonstrate that the proposed methods outperform
 existing detection and identification methods  {\color{black} on a synthetic dataset and for line outage identification in power systems from the IEEE 118-bus test case}.
\end{abstract}
\section{Introduction}
Graph signals arise in a wide range of applications, ranging from physical networks, such as power,  transportation, and communication systems, to data-driven networks, where the graph is used to represent relations among the observed data, such as in social networks. The emerging field of graph signal processing (GSP) extends classical signal processing methodologies, such as filtering and sampling, to the graph domain \cite{Aliaksei,Shuman,Antonio,Milos,Tirza_GSP}. 
Based on this GSP theory, different graph signals are represented as white noise that has been filtered on the graph topology by a graph filter.
Most GSP methods are based on the assumption that the underlying network topology is known. However,  even if the original topology of the network is known, small changes in this topology may occur  that significantly degrade the performance of GSP tasks \cite{Ceci}.  In particular, edge disconnections, i.e.  links between the graph vertices that have been dropped, is a common problem, especially in physical networks. For example, in electrical networks, the problem of identifying line outages that happen due to environmental factors,  damages, aging, and malicious attacks, is a significant problem \cite{Goldsmith,Giannakis,Wu,Tate_Overbye2008}. Additional examples are identifying traffic congestion and blockages in transportation networks \cite{Kwan}, and detecting links that  drop %
in wireless communication networks because of random blocking or fading  \cite{Yoon}. Accordingly, a fundamental question 
is how to use measured graph signals to identify edge disconnections, where the original underlying network structure is known. 

Numerous works in the literature have focused on (full) graph-topology learning. A popular approach in the context of graphical models is the graphical Lasso, which is based on relating between the inverse of the sample covariance matrix, i.e. the precision matrix, and the connectivity of the graph \cite{Dempster,Friedman,Ami}. 
Recent GSP-based network inference frameworks are based on recovering  a graph shift operator (GSO),  which encodes direct relationships between
the signal elements from observed indirect relationships
generated by graph diffusion
processes  \cite{Segarra,Pasdeloup,Thanou,Shafipour}. Other works consider
inference of dynamic networks and nonlinear models \cite{giannakis1,Ariel,Ariel2}. 
However, the full topology identification methods   are inefficient (in the sense of the amount of data required) and suboptimal (in terms of performance) for detecting a few specific topology disconnections, where the nominal topology is known.

Detection of graph topology changes  with different prior knowledge on the topology has been investigated.
The problem of detecting changes in a sequence of graphs by spectral methods has been considered in \cite{xxx,yyy}.
Anomalous subgraph detection deals with detecting small, anomalous subgraphs embedded into background networks with known properties (such as  the modularity matrix  eigenvectors  \cite{subgraph1,subgraph4}). However, the methods in \cite{xxx,yyy,subgraph1,subgraph4} are based on data in the form of random graphs and do not consider data measured on the vertices of a graph. Recent works discuss
detecting changes in graph topology from graph signals \cite{Matched,Elvin,subgraph5}. In \cite{Matched,Elvin},  matched subspace detectors have been developed to decide 
which one of two given graphs match better with a given 
 dataset, under  the assumptions of bandlimitedness in the graph Fourier domain but 
 without making any assumption on the nature of the change itself.
 Using information regarding the nature of the change in the topology, as considered in this paper, is expected to improve the detection performance. 
Several 
 edge exclusion tests  have been proposed  for general graphical models by using  the partial spectral coherence \cite{graphical1},   Matsuda test statistic \cite{graphical2}, and   a generalized likelihood ratio test (GLRT) \cite{GGM,graphical3} for detecting edge exclusion.  
In \cite{Eduardo},  a new algorithm is proposed for learning the Laplacian  matrix of the graph in Gaussian Markov random field (GMRF) models,  where the  connectivity is assumed to be known. The simulations  in \cite{Eduardo} show that this algorithm can be used to identify connectivity mismatches.

In this paper, we consider the problem of detecting disconnections in the topology based on graph signals that are represented as the output  of smooth graph filters \cite{Vassilis,Egilmez,Dong_Bronstein_Frossard2020}. We formulate the hypothesis testing problem for deciding the status (connected/disconnected) of a specific subset of edges. We develop the likelihood ratio test (LRT) for this hypothesis testing and show that the LRT is a function of  the graph-frequency energy levels in the graph spectral domain. In addition, for the noiseless GMRF model, the LRT  is shown to be  a   graph-smoothness detector, which can be implemented based solely on local measurements. 
Next,  we develop the  maximum likelihood (ML) decision rule for the
 M-ary hypothesis testing problem of identifying all the combinations of edge disconnections in the network.
 We show that 
the ML decision rule  consists of parallel LRTs of the binary hypothesis testing problems of each subset of edges. 
 However,  the ML decision rule leads to an exhaustive search over the candidate removed edges, which requires testing a number of hypotheses that is exponential in 
 the number of edges and is  practically infeasible.
Thus,
we propose a new low-complexity greedy approach for the identification problem that identifies  a single disconnected edge at each iteration.
We then propose a neighboring strategy that reduces 
the computational complexity of the
greedy approach  even further by 
implementing the test locally,
 based solely  on the measurements at neighboring vertices.
 Finally, we perform numerical simulations {\color{black}on synthetic data and for outage identification in a power system,} showing that the proposed approaches can efficiently identify the disconnected edges  and outperform existing approaches.

The rest of the paper is organized as follows. In Section \ref{Representation Model of Network Data}, we present a background on GSP and the observation model. In Section \ref{Detection of Topology Change}, we develop the LRT and its properties for detecting specific edges in the network. In Section \ref{classification} and Section \ref{Proposed Solutions}, we present the ML decision rule and the low-complexity greedy approaches, respectively. In Section \ref{Simulations}, a simulation study is presented. Finally, the paper ends with conclusions in Section \ref{Conclusion}.

In the rest of this paper, boldface lowercase letters denote vectors, and boldface uppercase letters denote matrices. The operators $(\cdot)^{T}$, $(\cdot)^{-1}$, $(\cdot)^{\dagger}$, $|\cdot|_+$, and  $\text{Tr}(\cdot)$   denote the transpose, inverse, Moore-Penrose pseudo-inverse, pseudo-determinant (i.e. the product of the non-zero eigenvalues of a positive semi-definite (PSD) matrix), and trace, respectively. 
The matrix ${\text{diag}}(\avec)$ is a diagonal matrix, whose diagonal elements are given by $\avec$. 
The $i$-th element of the vector $\avec$  is denoted by $a_i$.
The sub-matrix of the matrix $\Amat$ with the rows and columns indicated by the indices subsets $\mathcal{S}_r$ and  $\mathcal{S}_c$, respectively, is denoted by
 $\Amat_{\mathcal{S}_r,\mathcal{S}_c}$ 
For simplicity, we denote $\Amat_{\mathcal{S},\mathcal{S}}$ by $\Amat_{\mathcal{S}}$.
 In particular, $A_{i,j}$ is the $(i,j)$th entry of $\Amat$. The vectors $\onevec$ and $\zerovec$  are column vectors of ones and zeros, respectively, with appropriate dimensions. The matrices  $\Imat$ and $\Omat$ are the identity matrix and the zero matrix, respectively, with appropriate dimensions, where the $i$-th column of $\Imat$ is denoted by $\evec_i$.
 For a symmetric  matrix,
	$\Amat$, $\Amat\succeq\Omat$ means that $\Amat$ is a PSD matrix.
The notations ${\mathcal{O}}(\cdot)$ and the $\Omega(\cdot)$ are the commonly-used notations that describe the complexity \cite{complexity_book}.
\section{Measurement model: Output of a GSP filter}\label{Representation Model of Network Data}
In this section, we  present the background for GSP in Subsection \ref{Graph Signal Processing Overview} and define the smooth graph filter in Subsection \ref{Smooth graph filter}. Then, we describe the considered measurement model as an output of a smooth GSP filter in Subsection \ref{Measurements Model}. 

\subsection{Background: Graph Signal Processing (GSP)} \label{Graph Signal Processing Overview}
Consider an undirected, connected, weighted graph $\mathcal{G}=(\mathcal{V}, \mathcal{E}, \Wmat)$, where $\mathcal{V}$ and $\mathcal{E}$ are sets of vertices and edges, respectively. The matrix $\Wmat\in \mathbb{R}^{N \times N} $ is a weighted adjacency matrix of the graph, where $N\triangleq|\mathcal{V}|$ is the number of vertices in the graph. If there is an edge $(i,j)\in \mathcal{E}$ connecting vertices $i$ and $j$, the entry $W_{i,j}>0$ represents the weight of the edge; otherwise, $W_{i,j}=0$, \cite{Shuman,Antonio}. A common way to represent the graph topology is by the Laplacian matrix, defined by
 \begin{align*}\label{1}
     \Lmat\triangleq\text{diag}(\Wmat\onevec)-\Wmat. \numberthis
 \end{align*}
The Laplacian matrix is a real PSD matrix. Hence, its associated
 eigenvalue decomposition (EVD)   is given by
 \begin{equation}\label{eigendecomposing}
    \Lmat=\Umat^{(\Lmat)}\Lambdamat^{(\Lmat)} \big(\Umat^{(\Lmat)}\big)^{T},
\end{equation}
where the columns of $\Umat^{(\Lmat)}\:\in \mathbb{R}^{N \times N}$ are the eigenvectors of $\Lmat$ and $(\Umat^{(\Lmat)})^{-1}=(\Umat^{(\Lmat)})^T$. The matrix $\Lambdamat^{(\Lmat)}\in \mathbb{R}^{N \times N}$ is a diagonal matrix consisting of the eigenvalues of $\Lmat$, such that $0=\lambda^{(\Lmat)}_1<\lambda^{(\Lmat)}_2 \leq \ldots \leq \lambda^{(\Lmat)}_{N}$. 
For the sake of simplicity of presentation, the analysis and discussions in this paper are  under the assumption that the graphs  are  connected graphs, i.e. they have  a single zero eigenvalue \cite{Newman}. 

A graph signal $\avec:\mathcal{V}\rightarrow\mathbb{R}^{N}$ is an $N$-dimensional vector measured over the vertices of the graph. The graph Fourier transform (GFT) of the graph signal $\avec$ is defined as \cite{Shuman,Aliaksei}:
  \begin{equation}\label{e:GFT}
      \tilde{\avec}^{(\Lmat)}=\big(\Umat^{(\Lmat)}\big)^{T}\avec.
  \end{equation}
 Similarly, the inverse GFT (IGFT) is obtained by left multiplication of $\tilde{\avec}^{(\Lmat)}$ by  $\Umat^{(\Lmat)}$. The graph Laplacian quadratic form, also named the Dirichlet energy, is defined as
\begin{align*}\label{global smooth}
    Q_{\Lmat}(\avec) &\triangleq \avec^{T}\Lmat\avec= \frac{1}{2}\sum_{i\in \mathcal{V}}\sum_{j\in \mathcal{N}_i}{W_{i,j}(a_i-a_j)^2}\\
    &= \sum_{(i,j)\in \mathcal{E}}{W_{i,j}(a_i-a_j)^2},\numberthis
\end{align*} 
where $\mathcal{N}_i$ is the set of vertices connected to vertex $i$ by an edge in $\mathcal{E}$, and the first equality in (\ref{global smooth}) is obtained by substituting (\ref{1}). A smooth graph signals are signals with ``small" Dirichlet energy in (\ref{global smooth}). Intuitively, since the weights are non-negative, a smooth graph signal is considered to be a good match with the graph if the signal values are close to their neighbors' values \cite{Shuman}, as described on the right-hand side (r.h.s.) of (\ref{global smooth}). 
Substitution of (\ref{eigendecomposing}) and (\ref{e:GFT}) in (\ref{global smooth}) results in 
\begin{align*}\label{Parseval}
    Q_{\Lmat}(\avec) &= (\tilde{\avec}^{(\Lmat)})^{T}\Lambdamat^{(\Lmat)}\tilde{\avec}^{(\Lmat)}=\sum_{n=1}^{N}\lambda^{(\Lmat)}_n\big(\tilde{a}^{(\Lmat)}_n\big)^2\numberthis{},
\end{align*}
where $\tilde{a}^{(\Lmat)}_n$ is the $n$th element of the GFT defined in \eqref{e:GFT},  $n=1,\ldots,N$.
Therefore, a smooth graph signal with a small $Q_{\Lmat}(\avec)$ is associated with GFT coefficients that have a decaying behavior \cite{Shuman,Graph1}. 

\subsection{Smooth graph filter}\label{Smooth graph filter}
Some of these graph signals may be represented as white noise that has been filtered on the graph topology by a graph filter.
A graph filter is defined as follows \cite{Shuman}:
\begin{align*} \label{notation}
    h(\Lmat)&\triangleq\Umat^{(\Lmat)}h(\Lambdamat^{(\Lmat)})\big(\Umat^{(\Lmat)}\big)^T\\
    &=\Umat^{(\Lmat)} \text{diag}\big(h(\lambda^{(\Lmat)}_1),\ldots,h(\lambda^{(\Lmat)}_{N})\big)\big(\Umat^{(\Lmat)}\big)^T\numberthis,
\end{align*}
where $h(\cdot)$  is the transfer function of the filter.
A graph filter  is a linear operator applied on an input graph signal, $\avec_{\text{in}}$, such that the output graph signal, $\avec_{\text{out}}$, 
 satisfies \cite{Shuman,Egilmez,Moura}
\begin{equation}\label{9}
   \avec_{\text{out}}=h(\Lmat)\avec_{\text{in}}.
\end{equation} 
In this paper, we consider smooth graph filters.
Table \ref{FT} presents smooth graph filters that are commonly used in the GSP and network science literature. 
\begin{table}[hbt]
\centering
\begin{tabularx}{0.5\textwidth}{|X|c|X| }
\hline
 Graph Filter &  $h(\lambda)$ & {\color{black}Covariance matrix of $\avec_{out}$ for $Cov(\avec_{in})=\Imat$  }\\ 
 \hline
\vspace{-0.5cm} GMRF with a Laplacian precision matrix \cite{Dong,Vassilis} & {\parbox{0.15\textwidth}{\begin{align}\label{GMRF filter}
 \begin{cases}\frac{1}{\sqrt{\lambda}} \quad &\lambda\neq0\\
    0 \quad& \lambda=0\end{cases}\end{align}}}& {\parbox{0.1\textwidth}{\begin{align*}
{\color{black}\Lmat^\dagger}\end{align*}}} \\
  \hline 
  Regularized Laplacian (Tikhonov) \cite{Vassilis,Zhu} &  {\parbox{0.15\textwidth}{\begin{align}\label{Tikhonov filter}\frac{1}{1+\alpha\lambda},~\alpha>0 \end{align}}}& {\parbox{0.1\textwidth}{\begin{align*}{\color{black}
(\Imat+\alpha\Lmat)^{-2}}\end{align*}}}   \\
  \hline
  Heat Diffusion Kernel \cite{Thanou,Zhu,Vassilis} & {\parbox{0.15\textwidth}{\begin{align}\label{diffusion filter}\hspace{-0.32cm}\exp{(-\tau\lambda}),~\tau>0\end{align}}}& 
  {\parbox{0.1\textwidth}{\begin{align*}{\color{black}
\exp{(-2\tau\Lmat})}\end{align*}}} \\
  \hline 
\end{tabularx}
 \caption{Examples of smooth graph filters.}
 \label{FT}
\end{table}

\hspace{-0.4cm}In the following, we give a formal definition of smooth graph filters, which states that the expected Dirichlet energy  of the output graph signal is lower than that  of the input graph signal.
\begin{mydef}\label{smoothdef}
Let the elements of the input graph signal, $\avec_{\text{in}}$, be independent and identically distributed (i.i.d.)  
 zero-mean Gaussian random variables. Then, $h(\cdot)$ from  \eqref{notation} is
a smooth graph filter if 
 \begin{align*}\label{SGF}
     \frac{{\rm{E}}[Q_{\Lmat}(\avec_{\text{out}})]}{{\rm{E}}[Q_{\Lmat}(\avec_{\text{in}})]}<1,\numberthis
 \end{align*}
 where the Dirichlet energy, $Q_{\Lmat}(\cdot)$, is defined in (\ref{global smooth})
 and  $\avec_{\text{out}}$ is given in \eqref{9}.
\end{mydef}
By using (\ref{global smooth}), it can be verified that smoothness according to Definition \ref{smoothdef} is satisfied for the graph filters in (\ref{Tikhonov filter}) and (\ref{diffusion filter}).
In addition, 
for the GMRF graph filter from (\ref{GMRF filter}), 
\eqref{SGF} holds
under the assumption that the Laplacian matrix satisfies $\text{Tr}(\Lmat)>1$. Smooth graph signal corresponds to slow variations within the neighboring/connected vertices \cite{Shuman,dakovic}.
Similarly, it can be shown that these filters are low-pass graph filters \cite{Anna1,lital}.

\subsection{Measurement model}\label{Measurements Model}
In this paper, we consider a graph signal model as an output of a smooth graph filter, $h(\Lmat)$, in the form given in (\ref{9}), where the input graph signal is   white Gaussian noise. This model is presented in \cite{Vassilis,Egilmez,Dong_Bronstein_Frossard2020} and  it is shown that many signals over networks could be represented by this model, e.g. in power systems \cite{2021Anna,Drayer_2020,lital}. Thus, the measurement model is given by
\begin{align}\label{model}
    \yvec[m]= h(\Lmat)\xvec[m] +\wvec[m], ~ m = 1 \ldots M,
\end{align}
where $m$ is a time index, $\{\xvec[m]\}_{m=1}^M$,  is a sequence of  i.i.d.  Gaussian random vectors with zero mean and a diagonal covariance matrix, $\sigma^2_{\xvec}\Imat$,  i.e. $\xvec[m] \sim \mathcal{N}(\zerovec,\sigma^2_{\xvec}\Imat)$, $m=1,\ldots,M$, where $\sigma^2_{\xvec}>0$ is assumed to be known.  The graph filter, $h(\cdot)$, is assumed to be known. 
The noise $\{\wvec[m]\}_{m=1}^M$, is a sequence of i.i.d. zero–mean, Gaussian random vectors with  a covariance matrix $\sigma^2_{\wvec}\Imat$, where $\sigma^2_{\wvec}>0$ is known, i.e. $\wvec[m] \sim \mathcal{N}(\zerovec,\sigma^2_{\wvec}\Imat)$, $m=1 \ldots M$. It is also assumed that the sequences $\{\xvec[m]\}_{m=1}^M$ and $\{\wvec[m]\}_{m=1}^M$ are mutually independent.

Under these assumptions,  the output graph signal, $\yvec[m]$, $m=1,\ldots,M$, from (\ref{model}), is a sequence of i.i.d. zero–mean, Gaussian random vectors with the covariance matrix 
\begin{eqnarray}\label{Sigma}
    \bsigma(\Lmat)&\triangleq& \sigma^2_{\xvec}h^2(\Lmat)+\sigma^2_{\wvec}\Imat
    \nonumber\\&=&\sigma^2_{\xvec}\Umat^{(\Lmat)}h^2(\Lambdamat^{(\Lmat)})\big(\Umat^{(\Lmat)}\big)^T+\sigma^2_{\wvec}\Imat,
\end{eqnarray}
where we use the fact that $ h(\Lmat)h^T(\Lmat)=h^2(\Lmat)$, since $\Lmat$ is a symmetric matrix, and the last equality stems from \eqref{notation}. As a result, the log-likelihood of the augmented output vector of $M$ time samples, $\yvec\triangleq[\yvec^T[1],\dots,\yvec^T[M]]^{T}$, is 
\begin{align*} \label{probability 1}
     \log f(\yvec;\Lmat)=&-\frac{MN}{2}\log(2\pi)-\frac{M}{2}\log\big(|\sigma^2_{\xvec}h^2(\Lmat)+\sigma^2_{\wvec}\Imat|_+\big)\\ &-\frac{M}{2}\text{Tr}\bigg(\big(\sigma^2_{\xvec}h^2(\Lmat)+\sigma^2_{\wvec}\Imat \big)^{\dagger}\Smat_{\yvec}\bigg),\numberthis
 \end{align*}
where the sample covariance matrix is given by
\begin{equation} \label{Sample covariance}
    \Smat_{\yvec}\triangleq \frac{1}{M}\sum_{m=1}^M\yvec[m]\yvec^{T}[m].
\end{equation} 

The use of the pseudo-inverse and the pseudo-determinant in the log-likelihood  in (\ref{probability 1}) is since the covariance matrix in (\ref{Sigma}) may be a singular matrix. For example, for the GMRF graph filter in (\ref{GMRF filter}) from Table \ref{FT}  with $\sigma_{\wvec}^2=0$,   the matrix in (\ref{Sigma}) is reduced to
\begin{align*}\label{GMRF_cov}
    \bsigma(\Lmat)=\sigma_{\xvec}^2h^2(\Lmat)+\sigma^2_{\wvec}\Imat=\sigma_{\xvec}^2\Lmat^{\dagger},\numberthis
\end{align*}
 which is a singular matrix. More details about the use of a singular covariance matrix for the Gaussian distribution can be found in \cite{det,Andrew}.

It can be seen from (\ref{Sigma}) that in the considered  model the graph topology determines the covariance matrix of the output graph signal. In fact, the graph filter, $h(\Lmat)$, ``colors" the input graph signal using the network connectivity.
Therefore, when there are edge disconnections it affects the covariance matrix for the graph signal. 
In this work, we assume that the measurements are obtained from the model in (\ref{model}) and our goal is to detect (Section \ref{Detection of Topology Change}) and localize (Section \ref{classification}) edge disconnection based on this model.

\section{Detection of a topology change}\label{Detection of Topology Change}
In this section, we  develop the LRT for detecting the status (connected/disconnected) of a specific set of edges, based on the graph filter output model described in Subsection \ref{Measurements Model}. In this section, we assume that the exact topology of the graph is known under both hypotheses. In Subsection \ref{Problem Formulation} and Subsection \ref{LRT_section}, we present the hypothesis testing of this problem and develop the appropriate LRT, respectively. In Subsection \ref{GSP interpretation},  we present the LRT in the graph spectral domain.
Then, in Subsection \ref{A Gaussian Graphical Model interpretation}, we develop
the LRT for the special case of a GMRF graph filter.

\subsection{Problem formulation: Detection}\label{Problem Formulation}
We consider two candidate graphs, $\mathcal{G}^{(0)} = (\mathcal{V}, \mathcal{E}^{(0)},\Wmat^{(0)} )$ and
$\mathcal{G}^{(k)} = (\mathcal{V}, \mathcal{E}^{(k)},\Wmat^{(k)} )$, that are associated with the Laplacian matrices $\Lmat^{(0)}$ and $\Lmat^{(k)}$, respectively.
In fact, in this section the index $k$ can be treated as $k=1$, and the use of a general index $k$ is only for the sake of  simplicity of presentation for the case of  M-ary hypothesis testing in the following sections.  
The graph $\mathcal{G}^{(0)}= (\mathcal{V}, \mathcal{E}^{(0)},\Wmat^{(0)} )$ is the initial graph model of the network, representing the normal condition scenario with the associated known Laplacian matrix, $\Lmat^{(0)}$.
The graph $\mathcal{G}^{(k)}$ is a subgraph of $\mathcal{G}^{(0)}$,  which is obtained by removing a series of edges, $\mathcal{C}^{(k)}$, from the edge set, $ \mathcal{E}^{(0)}$.
Thus, the set of edges satisfies  
$\mathcal{E}^{(k)}=\mathcal{E}^{(0)}\setminus\mathcal{C}^{(k)}$.
The matrix $\Emat^{(i,j)}$, which corresponds to a single-edge disconnection of the edge $(i,j)\in \mathcal{E}^{(0)}$, is defined as 
\begin{equation} \label{topology error}
    \Emat^{(i,j)}\triangleq L^{(0)}_{i,j}(\evec_{i}\evec_{j}^{T}+\evec_{j}\evec_{i}^{T}-\evec_{j}\evec_{j}^{T}-\evec_{i}\evec_{i}^{T}).
\end{equation}
Accordingly, the Laplacian matrix, $\Lmat^{(k)}$, can be written as the sum of single-edge disconnections as follows:
\begin{align*}\label{multierror}
    \Lmat^{(k)}&\triangleq\Lmat^{(0)}-\Emat^{(k)},\numberthis{}
    \end{align*}
where
\begin{align*}\label{multierror1}
\Emat^{(k)}&\triangleq\sum_{(i,j)\in {\mathcal{C}}^{(k)}  }\Emat^{(i,j)} \numberthis{}.
\end{align*}
 It can be verified that $\Lmat^{(k)}$ from (\ref{multierror})  is a valid Laplacian matrix for any subset ${\mathcal{C}}^{(k)}\subset\mathcal{E}^{(0)}$.

Based on the measurement model in (\ref{model}), the  detection problem between the two topologies can be formulated as the following binary hypothesis testing problem:
\begin{align*}\label{hypothesis testing}
    {\mathcal{H}}_0: \quad& \yvec[m]= h(\Lmat^{(0)})\xvec[m] +\wvec[m] \\
    {\mathcal{H}}_1: \quad& \yvec[m]= h(\Lmat^{(k)})\xvec[m] +\wvec[m], \quad m = 1 \ldots M \numberthis{}.
\end{align*}
That is, the graph signal under each hypothesis is an output of a different topology, where $\Lmat^{(0)}$ represents the initial graph  and $\Lmat^{(k)}$ is  the Laplacian matrix from  (\ref{multierror}), which is obtained by a series of  disconnections of the edges in ${\mathcal{C}}^{(k)}$. The graph filter, $h(\Lmat)$, the noise, and the input graph signal statistics for the two hypotheses are as described in Subsection \ref{Measurements Model}. 
\subsection{LRT}\label{LRT_section}
The hypothesis testing problem in (\ref{hypothesis testing}), which assumes that $h(\cdot)$, $\Lmat^{(0)}$,  $\Lmat^{(k)}$, and the statistics of the noise and the input signal, are all known, is equivalent to the problem of testing the structured covariance matrix of random Gaussian vectors, which has been studied extensively in the literature (see, e.g. \cite{Kay,cov1,Ramirez,Soloveychik1}). The LRT  for this case is given by
\begin{equation}\label{LRT_def}
     \log\Big(\frac{f(\yvec;\Lmat^{(k)})}{f(\yvec;\Lmat^{(0)})}\Big)\lessgtr^{\mathcal{H}_0}_{\mathcal{H}_1}\gamma,
\end{equation}
where $\log f(\yvec;\Lmat^{(0)})$ and $\log f(\yvec;\Lmat^{(k)})$ are the log-likelihoods under the hypotheses ${\mathcal{H}}_0$ and ${\mathcal{H}}_1$, respectively, that are obtained by substituting $\Lmat=\Lmat^{(0)}$ and $\Lmat=\Lmat^{(k)}$ in (\ref{probability 1}). Hence, by substituting (\ref{probability 1}) under each hypothesis in (\ref{LRT_def}), the LRT associated with the Laplacian $\Lmat^{(k)}$ is given by
\begin{align*}\label{LRT1}
    l(\yvec|\Lmat^{(k)})\lessgtr^{\mathcal{H}_0}_{\mathcal{H}_1}\gamma' \numberthis,
\end{align*}
where 
\beqna\label{LRT11}
    l(\yvec|\Lmat^{(k)})&\triangleq&
    \text{Tr}\bigg(\big(\sigma^2_{\xvec}h^2(\Lmat^{(0)})+\sigma^2_{\wvec}\Imat\big)^{\dagger}\Smat_{\yvec}\bigg)
        \nonumber\\&&
    -\text{Tr}\bigg(\big(\sigma^2_{\xvec}h^2(\Lmat^{(k)})+\sigma^2_{\wvec}\Imat\big)^{\dagger}\Smat_{\yvec}\bigg),
\eeqna
and the threshold is $\gamma'=\gamma+\rho(\Lmat^{(k)})$, in which
\begin{align*}\label{rho p}
    \rho(\Lmat^{(k)})\triangleq\log\bigg(\frac{|\sigma_{\xvec}^2h^2(\Lmat^{(k)})+\sigma^2_{\wvec}\Imat|_+}{|\sigma_{\xvec}^2h^2(\Lmat^{(0)})+\sigma^2_{\wvec}\Imat|_+}\bigg).\numberthis
\end{align*} 
 Thus, the sufficient statistics of the LRT in the time domain is the sample covariance matrix, $\Smat_{\yvec}$.
\subsection{Interpretation of the LRT in the graph spectral domain}\label{GSP interpretation}
In this subsection, we present the GSP form of the LRT  from (\ref{LRT1}) based on its projection onto the graph spectral domain. 
For the sake of simplicity, 
we assume  that  
 $\bsigma(\Lmat^{(k)})$ and  $\bsigma(\Lmat^{(0)})$, as defined in
(\ref{Sigma}), are non-singular matrices. Then, by  applying the matrix inversion lemma (see, e.g. Eq. (1) in\cite{inverse}) on the r.h.s. of (\ref{Sigma}) after substituting  $\Lmat=\Lmat^{(k)}$, we obtain that
the inverse of the  covariance matrix 
 satisfies
\begin{align*}
\label{inverse sigma}
  \big( \bsigma(\Lmat^{(k)})\big)^{-1}= \big(\sigma^2_{\xvec}h^2(\Lmat^{(k)})+\sigma^2_{\wvec}\Imat\big)^{-1}\hspace{2cm}
\\=
   \frac{1}{\sigma^2_{\wvec}}\Imat-\frac{\sigma^2_{\xvec}}{\sigma^2_{\wvec}}\Umat^{(\Lmat^{(k)})} h(\Lambdamat^{(\Lmat^{(k)})})\hspace{2.75cm}\\
   \times(\sigma^2_{\wvec}\Imat+\sigma^2_{\xvec}h^2(\Lambdamat^{(\Lmat^{(k)})}))^{-1} h(\Lambdamat^{(\Lmat^{(k)})})\big(\Umat^{(\Lmat^{(k)})}\big)^T. \numberthis
\end{align*}
Moreover, by using the GFT definition in (\ref{e:GFT}), we obtain that the GFT of the output graph signal at time $m$ with respect to (w.r.t.) the Laplacian matrix $\Lmat^{(k)}$ is
\be\label{yhat}
    \tilde{\yvec}^{(\Lmat^{(k)})}[m]=(\Umat^{(\Lmat^{(k)})})^T\yvec[m],~m=1,\ldots,M.
\ee
Thus, the $n$th element of the mean-squared  GFT of the output graph signal, 
is defined as follows: 
\begin{align*}\label{cofficisent}
    \psi^{(\Lmat^{(k)})}_{n}\triangleq \frac{1}{M}\sum_{m=1}^M\big(\big[\tilde{\yvec}^{(\Lmat^{(k)})}[m]\big]_n\big)^2 \quad n=1,\ldots,N\numberthis{}.
\end{align*}
The expression in (\ref{cofficisent}) can be interpreted  as the $n$th \textit{graph-frequency energy level}.
 By substituting (\ref{Sample covariance}), (\ref{inverse sigma}), and (\ref{cofficisent}) for  general $k$  and for $k=0$ in (\ref{LRT11}), the LRT can be rewritten as
\begin{align*}\label{LRT_GsP}
    l(\yvec|\Lmat^{(k)})= \frac{\sigma_{\xvec}^2}{\sigma_{\wvec}^2}&\bigg(\sum_{n=1}^N\frac{h^2(\lambda_n^{(\Lmat^{(k)})})}{\sigma_{\wvec}^2+\sigma_{\xvec}^2h^2(\lambda_n^{(\Lmat^{(k)})})}\psi^{(\Lmat^{(k)})}_{n}\\
    &-\frac{h^2(\lambda_n^{(\Lmat^{(0)})})}{\sigma_{\wvec}^2+\sigma_{\xvec}^2h^2(\lambda_n^{(\Lmat^{(0)})})}\psi^{(\Lmat^{(0)})}_{n}\bigg),\numberthis
\end{align*}
where, since $h(\Lambdamat^{(\Lmat)})$ is a diagonal matrix, we have
\beqna
\label{diag_mat}
\left[h(\Lambdamat^{(\Lmat)})
    (\sigma^2_{\wvec}\Imat+\sigma^2_{\xvec}h^2(\Lambdamat^{(\Lmat)}))^{-1} h(\Lambdamat^{(\Lmat)})\right]_{i,j}\hspace{1.25cm}
    \nonumber\\=\left\{\begin{array}{lr}
    \frac{h^2(\lambda_j^{(\Lmat)})}{\sigma_{\wvec}^2+\sigma_{\xvec}^2h^2(\lambda_j^{(\Lmat)})},&{\text{if }} i=j\\
    0,&{\text{otherwise}}
    \end{array}\right.
    \eeqna
     for any Laplacian matrix, $\Lmat$.
It can be seen from (\ref{LRT_GsP}) that the sufficient statistic for the LRT  is the graph-frequency energy levels, $\{\psi^{(\Lmat^{(0)})}_{n},\psi^{(\Lmat^{(k)})}_{n}\}_{n=1}^N$. Moreover, for smooth graph filters, as defined in Definition \ref{smoothdef}, the weights of the graph-frequency energy levels given in \eqref{diag_mat},
$\frac{h^2(\lambda_n^{(\Lmat^{(0)})})}{\sigma_{\wvec}^2+\sigma_{\xvec}^2h^2(\lambda_n^{(\Lmat^{(0)})})}$ and $\frac{h^2(\lambda_n^{(\Lmat^{(k)})})}{\sigma_{\wvec}^2+\sigma_{\xvec}^2h^2(\lambda_n^{(\Lmat^{(k)})})}$,
tend to amplify low graph-frequencies.
Therefore, the LRT is governed by the low-graph frequencies, that can be associated with the local behaviour of the graph signals at neighboring vertices.

\subsection{LRT for the GMRF model with a Laplacian precision}\label{A Gaussian Graphical Model interpretation}
The Gaussian graphical model (GGM) is a prominent model that aims to handle large amounts of data collected in various applications and estimate the structure behind the data \cite{Rabbat}. Under this model, there is an exact correspondence between the non-zero entries in the precision matrix  and the existence of an edge between
the relevant vertices \cite{Dong}. In particular, the use of the  graph filter  in (\ref{GMRF filter}) leads to a GMRF model with the Laplacian as the precision matrix  \cite{Vassilis}.
In this subsection, we discuss the LRT for the special case of a GMRF  model.

Consider the GMRF in a noiseless scenario, $\sigma_{\wvec}^2=0$, then by substituting (\ref{Sample covariance}) and (\ref{GMRF_cov}) in (\ref{LRT11}) and using $(\Lmat^{\dagger})^{\dagger}=\Lmat$, we obtain that for this case
\begin{align*}\label{smooth1}
     l(\yvec|\Lmat^{(k)})=\frac{1}{\sigma_{\xvec}^2}\text{Tr}((\Lmat^{(0)}-\Lmat^{(k)})\Smat_{\yvec})\hspace{2cm}\\
    =\frac{1}{\sigma_{\xvec}^2M}\sum_{m=1}^M\big(Q_{\Lmat^{(0)}}(\yvec[m])-Q_{\Lmat^{(k)}}(\yvec[m])\big)\numberthis,
\end{align*}
where the Dirichlet energy, $Q_{\Lmat}(\cdot)$, is defined in (\ref{global smooth}). Thus, in this special case  the LRT  in (\ref{smooth1}) is a Dirichlet energy detector,
which measures if the measurements are smooth w.r.t. the Laplacian matrix $\Lmat^{(0)}$ or $\Lmat^{(k)}$ that are associated with graph $\mathcal{G}^{(0)}$ or graph $\mathcal{G}^{(k)}$. This result can be explained by the fact that the output graph signal energy  under the GMRF model tends to lie mainly in the low-frequency components \cite{Dong}.

\begin{figure}[hbt]
    \includegraphics[width=0.35\textwidth]{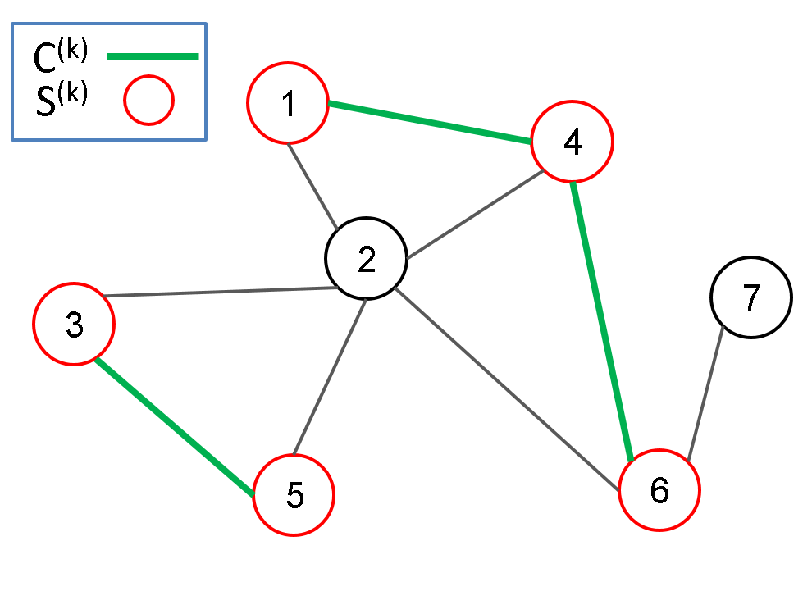}
    \caption{\textcolor{black}{Illustration of the subsets: A graph $\mathcal{G}^{(0)}=(\mathcal{V}, \mathcal{E}^{(0)}, \Wmat^{(0)})$ with $N = 7$ vertices and $|\mathcal{E}^{(0)}| = 9$ edges. 
    The graph $\mathcal{G}^{(k)}$ is  obtained by removing the series of edges, $\mathcal{C}^{(k)}=\{(1,4), (4,6), (3,5)\}$ that are associated with the
     subset of vertices, $\mathcal{S}^{(k)}=\{1,3,4,5,6\}$. If we assume that all the weights are equal to $1$, then, according to 
     \eqref{multierror1}, in this example\\
$\Emat^{(k)}=\sum_{(i,j)\in {\mathcal{C}}^{(k)}  }\Emat^{(i,j)}=\left[\begin{smallmatrix}
  1 &0&0&-1&0&0 &0\\
  0 &0&0&0&0&0 &0\\
  0 &0&1&0&-1&0 &0\\
  -1 &0&0&2&0&-1 &0\\
  0 &0&-1&0&1&0 &0\\
  0 &0&0&-1&0&1 &0\\
  0 &0&0&0&0&0 &0
\end{smallmatrix}\right]$.
     }}
    \label{fig:my_label1}
\end{figure}

Another observation regarding the LRT in  this case is as follows. By substituting \eqref{topology error}-\eqref{multierror1} in (\ref{smooth1}), we obtain
\beqna\label{e3}
 \hspace{-0.25cm}    l(\yvec|\Lmat^{(k)})
    =-\frac{1}{\sigma_{\xvec}^2M}\hspace{-0.15cm}\sum_{(i,j)\in\mathcal{C}^{(k)}}L_{i,j}^{(0)}
    \sum_{m=1}^M(y_i[m]-y_j[m])^2.
\eeqna
 Hence, it can be seen that the LRT, $l(\yvec|\Lmat^{(k)})$, in this case, includes only measurements that are measured at 
 the vertices that are associated with the edges in  $\mathcal{C}^{(k)}$, defined as follows.
\begin{mydef}
\label{subset_S}
The subset of the vertices that correspond to the edge set $\mathcal{C}^{(k)}$ is defined as
\begin{align*}
\label{subset_eq}
\mathcal{S}^{(k)}\define \{i  \in \mathcal{V}|  \exists j\in\mathcal{V} {\text{ s.t. }}
     (i,j)\in \mathcal{C}^{(k)}\} .\numberthis 
    \end{align*}
\end{mydef}
{\color{black}
In Fig. \ref{fig:my_label1} we illustrate  the notations used in this paper for the different subsets over an arbitrary graph.}

By using Definition \ref{subset_S}, 
 it can be verified that the elements of the matrix $ \Emat^{(k)}$ from  \eqref{multierror1} satisfy
\begin{align*}\label{E_sparse}
    \Emat^{(k)}_{m,l}=0, \text{ if }m\notin\mathcal{S}^{(k)}\text{ and/or  }l\notin\mathcal{S}^{(k)}.\numberthis
\end{align*}
Hence, by substituting (\ref{multierror}) in the first row of (\ref{smooth1}), we obtain
\begin{align*}\label{local_GMRF}
     l(\yvec|\Lmat^{(k)})=&\frac{1}{\sigma_{\xvec}^2}\text{Tr}\big(\Emat^{(k)}\Smat_{\yvec}\big)=\frac{1}{\sigma_{\xvec}^2}\sum_{i=1}^N\sum_{j=1}^N[\Emat^{(k)}]_{i,j}[\Smat_{\yvec}]_{j,i}\\
     =&\frac{1}{\sigma_{\xvec}^2}\sum_{i\in \mathcal{S}^{(k)}}\sum_{j\in \mathcal{S}^{(k)}}[\Emat^{(k)}]_{i,j}[\Smat_{\yvec}]_{j,i}\\
     =&\frac{1}{\sigma_{\xvec}^2}\text{Tr}\big([\Emat^{(k)}]_{\mathcal{S}^{(k)}}[\Smat_{\yvec}]_{\mathcal{S}^{(k)}}\big)\numberthis,
\end{align*}
where the third equality is obtained by substituting (\ref{E_sparse}) and the last equality is obtained by using the trace operator properties.
 From the observations in (\ref{e3}) and (\ref{local_GMRF}), it can be concluded that the measurements at all  vertices that do not belong to $\mathcal{S}^{(k)}$ do not affect the LRT and are non-informative for the hypothesis testing in \eqref{hypothesis testing}.
 This conclusion is aligned with the Hammersley-Clifford theorem \cite{hammer}, which states that a probability density function
(pdf) that satisfies  Markov properties w.r.t an undirected graph $\mathcal{G}$, such as   the GMRF model, can be factorized into positive functions defined on cliques that cover all the vertices and edges of the graph. As a result, the LRT for the GMRF model in (\ref{local_GMRF})  is the difference between two log-likelihood functions that   satisfy Markov properties and is only evaluated at the vertices that are in the subtraction of the cliques under hypothesis $\mathcal{H}_1$ from the cliques under $\mathcal{H}_0$, which are exactly the vertices in  $\mathcal{S}^{(k)}$.

\section{Identification of edge disconnections} \label{classification}
In this section, we investigate the problem of identifying general edge disconnections in a network that are not limited to a specific set of edges.
 In Subsection \ref{Problem Formulation1}, we formulate the M-ary hypothesis testing  of the edge disconnections identification problem. In Subsection  \ref{M-ary Maximum Likelihood Ratio Decision Rule}, we develop the ML decision rule for this hypothesis-testing problem. In Subsection \ref{GSP interpretation1}, we discuss the properties of the ML decision rule and its computational complexity.
\subsection{Problem formulation: 
 Identification }\label{Problem Formulation1}
We consider the following identification problem: the graph $\mathcal{G}^{(0)}= (\mathcal{V}, \mathcal{E}^{(0)},\Wmat^{(0)} )$ is the initial graph model of the network, representing the normal condition scenario with the associated known Laplacian matrix, $\Lmat^{(0)}$. Our  purpose is to identify the graph topology from a set of possible graphs,  $\big\{\mathcal{G}^{(k)} = (\mathcal{V}, \mathcal{E}^{(k)},\Wmat^{(k)} )\big\}_{k=0}^K$, where $\Lmat^{(k)}$ represents the Laplacian matrix of the corresponding graph $\mathcal{G}^{(k)}$. Each graph, $\mathcal{G}^{(k)}$,  $k=1,\ldots,K$, is a subgraph of $\mathcal{G}^{(0)}$ with  $\mathcal{E}^{(k)}=\mathcal{E}^{(0)}\setminus\mathcal{C}^{(k)}$, where
$\mathcal{C}^{(k)}$ is the $k$th set of edge  disconnections, as  described for the binary hypothesis testing in Subsection \ref{Problem Formulation}.

The identification problem between $K+1$ possible hypotheses  based on $M$ graph signals from (\ref{model}), $\yvec[m]$, $m=1,\ldots,M$, can be stated as  an M-ary hypothesis testing problem:
\begin{align*}\label{H}
    &{\mathcal{H}}_0: \quad \yvec[m]= h(\Lmat^{(0)})\xvec[m] +\wvec[m], \\
    &{\mathcal{H}}_k: \quad \yvec[m]= h(\Lmat^{(k)})\xvec[m] +\wvec[m],\quad m = 1 \ldots M,\numberthis
\end{align*}
for $k=1,\ldots,K$, where $\Lmat^{(k)}$ is defined in (\ref{multierror}). 
This problem is a simple multiple hypothesis testing problem, where all the parameters under each hypothesis are known. 
\subsection{ML decision rule}\label{M-ary Maximum Likelihood Ratio Decision Rule}
The ML decision rule maximizes the log-likelihood of the $K+1$ hypotheses in (\ref{H})  and is given by \cite{Kay}
\begin{equation}\label{def ML M array}
     \xi(\yvec)=\argmax_{0 \leq k \leq K}\log f(\yvec; \Lmat^{(k)}),
\end{equation}
where $\log f(\yvec;\Lmat^{(k)} )$ is the log-likelihood under hypothesis ${\mathcal{H}}_k$, which is obtained by substituting $\Lmat=\Lmat^{(k)}$ in (\ref{probability 1}), for $k=0,1,\ldots,K$. By dividing each of the $K+1$ log-likelihood functions on the r.h.s. of (\ref{def ML M array}) by the positive log-likelihood of the null hypothesis, $\log f(\yvec;\Lmat^{(0)})$,  (\ref{def ML M array}) can be rewritten as 
\begin{align*}\label{MML pro}
    \xi(\yvec)=\argmax_{0 \leq k \leq K}\bigg\{ \log\bigg(\frac{f(\yvec; \Lmat^{(k)}  )}{f(\yvec;\Lmat^{(0)})}\bigg)\bigg\}\numberthis{}.
\end{align*}
By substituting (\ref{probability 1}) for each likelihood function in  (\ref{MML pro})  and removing constant terms, the ML decision rule satisfies
    \begin{align*}\label{MML min1}
    \xi(\yvec)&=\argmax_{0 \leq k \leq K}l(\yvec|\Lmat^{(k)})-\rho(\Lmat^{(k)}), \numberthis
    \end{align*} 
where $l(\yvec|\Lmat^{(k)})$  
and $\rho(\Lmat^{(k)})$ are defined in 
 (\ref{LRT11}) and 
(\ref{rho p}), respectively.
It can be seen from (\ref{MML min1}) that under hypothesis ${\mathcal{H}}_0$ we obtain $l(\yvec|\Lmat^{(0)})-\rho(\Lmat^{(0)})=0$.

\begin{figure}[hbt]
    \centering
    \includegraphics[width=0.33\textwidth]{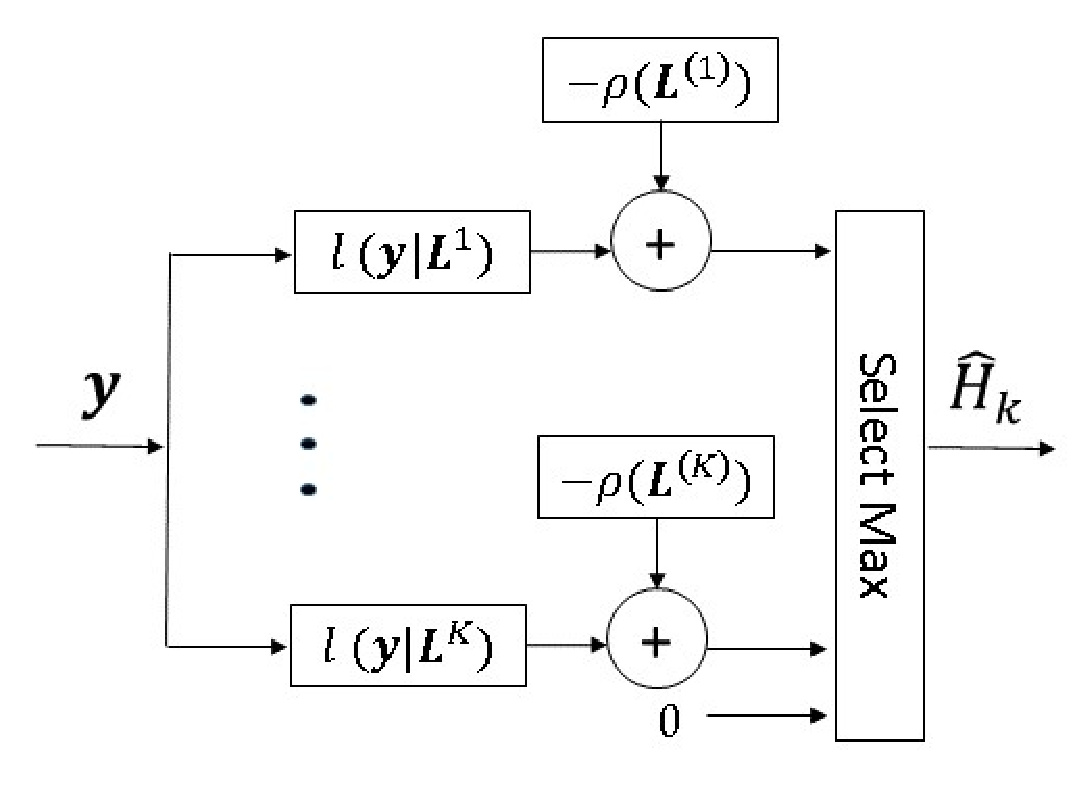}
    \caption{Block diagram of the ML decision rule:  $K$ LRTs  are implemented in parallel and summed with appropriate penalty functions, $-\rho(\Lmat^{(k)})$, $k=1,\ldots,K$. Then, the ML decision rule selects the hypothesis associated with the maximal values, where $0$ is  the value of the null hypothesis, $\mathcal{H}_0$. }
    \label{fig:MML}
\end{figure}

The term $l(\yvec|\Lmat^{(k)})$ is the  LRT  for the binary hypothesis testing between ${\mathcal{H}}_0$ and ${\mathcal{H}}_k$, i.e. detecting a specific edge disconnections set $\mathcal{C}^{(k)}$, as described in \eqref{hypothesis testing}.
Thus, the ML decision rule in (\ref{MML min1}) consists of two stages: first, implementing $K$ binary LRTs in parallel, where each LRT  is summed with an appropriate penalty function,  $-\rho(\Lmat^{(k)})$, which is independent of the measurement vector, $\yvec$. Second,
the ML decision rule selects the hypothesis associated with the maximal values, where $0$ is  the value of the null hypothesis, $\mathcal{H}_0$.
The ML decision rule is illustrated   in Fig. \ref{fig:MML}.
\vspace{-0.5cm}
\subsection{Remarks on the ML decision rule}\label{GSP interpretation1}
\subsubsection{Penalty function interpretation}
The data-independent term, $\rho(\Lmat^{(k)})$ in (\ref{rho p}),  can be interpreted as a penalty function on the considered topology change. In order to demonstrate this interpretation,  the following proposition describes the order relation between the penalty functions.
\begin{proposition}\label{eigenvalue}
Consider two connected graphs, $\mathcal{G}^{(k_1)}$ and $\mathcal{G}^{(k_2)}$, associated with the Laplacian matrices, $\Lmat^{(k_1)}$ and  $\Lmat^{(k_2)}$, respectively, and assume the following conditions:\\
C.1) The matrices $\bsigma(\Lmat^{(k_1)})$ and $\bsigma(\Lmat^{(k_2)})$, defined in
(\ref{Sigma}), are non-singular matrices.\\ C.2) $\mathcal{C}^{(k_2)}$ is a proper subset of $\mathcal{C}^{(k_1)}$, i.e.
$
\mathcal{C}^{(k_2)}\subset\mathcal{C}^{(k_1)}$.\\
 C.3) the graph filter, $h(\lambda)$, is a monotonic decreasing function of $\lambda$ for any $\lambda >0$.\\
Then, 
    \be \label{penalty pro}
   \rho(\Lmat^{(k_2)})\leq \rho(\Lmat^{(k_1)}).\numberthis
    \ee
\end{proposition}
\begin{IEEEproof} 
	The proof appears in Appendix \ref{B}.
\end{IEEEproof} 
It can be seen that the graph filters from (\ref{GMRF filter})-(\ref{diffusion filter}) in Table \ref{FT} are all monotonic decreasing functions for $\lambda>0$ and, thus, satisfy the conditions of Proposition \ref{eigenvalue}. 
According to this proposition, 
in the case of nested subsets
of edge disconnections,  the ML decision rule in (\ref{MML min1})  imposes a larger penalty on models with a larger number of edge disconnections, in order to avoid overfitting \cite{Stoica}.


\subsubsection{Interpretation  in the graph spectral domain}
Similar to in Subsection \ref{GSP interpretation}, the ML decision rule from (\ref{MML min1}) can be written in the graph spectral domain using the result in (\ref{LRT_GsP}) for the $k$th hypothesis. That is, the ML decision rule in (\ref{MML min1}) can be written as
\begin{align*}\label{MML GSP}
     \xi(\yvec)=&\argmax_{0 \leq k \leq K}
     \frac{\sigma_{\xvec}^2}{\sigma_{\wvec}^2}\bigg(\sum_{n=1}^N\frac{h^2(\lambda_n^{(\Lmat^{(k)})})}{\sigma_{\wvec}^2+\sigma_{\xvec}^2h^2(\lambda_n^{(\Lmat^{(k)})})}\psi^{(\Lmat^{(k)})}_{n}\\
    &-\frac{h^2(\lambda_n^{(\Lmat^{(0)})})}{\sigma_{\wvec}^2+\sigma_{\xvec}^2h^2(\lambda_n^{(\Lmat^{(0)})})}\psi^{(\Lmat^{(0)})}_{n}\bigg)-\rho(\Lmat^{(k)}),\numberthis{}
    \end{align*}
where $\rho(\Lmat^{(k)})$ is defined in (\ref{rho p}), and the $n$th graph-frequency energy level w.r.t. $\Lmat^{(k)}$, $\psi^{(\Lmat^{(k)})}_{n}$, is defined in (\ref{cofficisent}). 
It should be noted that the representation in the graph spectral domain emphasizes the fact that the  ML  decision  rule only requires  the evaluation of the $N K$ scalar parameters, $\psi^{(\Lmat^{(k)})}_{n}$, for any $k=0,\ldots,K$, $n=1,\ldots,N$.
Thus, in contrast with covariance matrix estimation problems, there is no need to assume that
the sample covariance matrix from \eqref{Sample covariance}, $\Smat_{\yvec}$, is a full rank matrix and we can achieve good identification performance even for a small number of samples, $M$.
This result is due to the fact that 
the M-ary hypothesis testing problem
in \eqref{H}  is a {\em{constrained}}  covariance matrix estimation problem with  a structured covariance matrix. 

\subsubsection{GMRF model with a Laplacian precision}
\label{A Gaussian Graphical Model interpretation1}
Similar to Subsection \ref{A Gaussian Graphical Model interpretation}, the following proposition states that  for the special case of a noiseless GMRF with a Laplacian precision matrix, the ML decision rule can be calculated locally for each candidate set $k=1,\ldots,K$, by observing only the related measurements in $\mathcal{S}^{(k)}$,  defined in Definition \ref{subset_S}.  
\begin{proposition}\label{pro_GMRF}
Consider a connected graph, $\mathcal{G}^{(k)}$, associated with the Laplacian matrix, $\Lmat^{(k)}$. Then, for  the noiseless GMRF filter in (\ref{GMRF filter}),    the term $l(\yvec|\Lmat^{(k)})-\rho(\Lmat^{(k)})$  on the r.h.s. of (\ref{MML min1}) is only a function of the measurements and  the second-order statistics that are  associated with the vertices in $\mathcal{S}^{(k)}$.
\end{proposition}
\begin{IEEEproof} 
It was shown in \eqref{local_GMRF} that  $l(\yvec|\Lmat^{(k)})$ is only a function of the measurements 
 that are measured at  vertices  in  $\mathcal{S}^{(k)}$. In addition,  in Appendix \ref{C} we prove that   under the assumptions of Proposition \ref{pro_GMRF}, $\rho(\Lmat^{(k)})$
 is only a function of the matrices $[\bsigma(\Lmat^{(k)})]_{\mathcal{S}^{(k)}}$ and 
 $[\bsigma(\Lmat^{(0)})]_{\mathcal{S}^{(k)}}$, i.e. of the second-order statistics of the graph signal  over the vertices in $\mathcal{S}^{(k)}$. Thus, the $k$th term  in \eqref{MML min1} depends only on the observation measured at the vertices in $\mathcal{S}^{(k)}$ and their statistics.
	\end{IEEEproof}

\subsubsection{Computational complexity}\label{Complexity of maximum likelihood test}
The ML decision rule in (\ref{MML min1}) is based on  evaluating the likelihood for each of the
 $K$ candidate hypotheses and then selecting the maximum among them. In the general worst-case scenario, where any of the edges can disconnect, we have
 \begin{align} \label{KK}
     K=\sum_{r=1}^{r_{\max}}{|\mathcal{E}^{(0)}| \choose r}, 
 \end{align}
where $r_{\max}$ denotes the considered maximum number of possible edge disconnections in the network.
Such a combinatorial optimization problem suffers from high computational complexity due to
the exhaustive search over the set of possibilities for each suspicious edge (connected/disconnected).
Moreover, the calculation of $l(\yvec|\Lmat^{(k)})$  
and $\rho(\Lmat^{(k)})$ in (\ref{LRT11}) and (\ref{rho p}), respectively, requires complexity of $N$-dimensional matrix inversion.
Thus, the computational complexity of the ML decision rule  grows exponentially with the graph size and it would be impractical for  practical networks.
In the next section, we develop  efficient low-complexity methods to deal with the edge disconnections identification problem.

\section{Greedy approaches for identifying edge disconnections }\label{Proposed Solutions}
In this section, we propose  low-complexity greedy approaches for edge disconnections identification.
We first present the basic greedy method in Subsection \ref{The Greedy approach and neighboring strategy}. Then, in Subsection \ref{Neighboring Strategy}, we present a modification: a neighboring strategy that is combined with the greedy method to reduce the computational complexity even further. Some remarks on these approaches are discussed  in Section \ref{Advantages}.
\subsection{Greedy approach}\label{The Greedy approach and neighboring strategy}
We assume a nominal topology, where
 only a small
percentage of edges may be disconnected, i.e.
$|\mathcal{C}^{(k)}|\ll|\mathcal{E}^{(0)}|$, $k=1,\ldots,K$, and the matrices $\Emat^{(k)}$, $k=1,\ldots,K$ are sparse. 
A
commonly-used heuristic for combinatorial  problems over graphs is a greedy algorithm \cite{Gomez}, which starts with an empty set
 and then, iteratively, in each step, adds the edge  which maximizes the objective.
In the considered identification problem, we propose a greedy approach, given in Algorithm \ref{greedy_al}, that starts with $\hat{\mathcal{C}}^0=\emptyset$ at the first ($l=0$) iteration.  Then,
at the $l$th iteration, we test all the available edges in the graph  and choose the edge that maximizes the marginal likelihood in (\ref{MML min1}) for a single edge: 
\begin{align*} \label{greedy_likelihood}
     \hat{k}=\argmax_{k=(i,j)\in \hat{\mathcal{E}}^{l}}  l(\yvec|\hat{\Lmat}^{l}-\Emat^{(k)})-\rho(\hat{\Lmat}^{l}-\Emat^{(k)}),\numberthis
   \end{align*} 
   where $\hat{\mathcal{E}}^{l}$ and $\hat{\Lmat}^{l}$ are the  edge set and the Laplacian matrix at the $l$th iteration that are initialized by $\hat{\mathcal{E}}^{0} = \mathcal{E}^{(0)}$ and $\hat{\Lmat}^{0}=\Lmat^{(0)}$, respectively, 
    where in the left terms, ``0" denotes the iteration index.
The functions $l(\cdot|\cdot)$ and $\rho(\cdot)$ in (\ref{greedy_likelihood}) are defined in (\ref{LRT11}) and (\ref{rho p}), respectively. 
  Afterwards, we test if the likelihood ratio of the chosen edge, $\hat{k}$, is higher than zero:  
\begin{align}\label{stopcondition}
        l(\yvec|\hat{\Lmat}^{l}-\Emat^{(\hat{k})})-\rho(\hat{\Lmat}^{l}-\Emat^{(\hat{k})})> 0. \numberthis
    \end{align}
 If \eqref{stopcondition} is satisfied, then: 1) we add $\hat{k}$ to the edge disconnections set, $\hat{\mathcal{C}}^{l}$; 2) we update 
 the edge set to $\hat{\mathcal{E}}^{l+1}$ by
 removing $\hat{k}$ from the current edge set, $\hat{\mathcal{E}}^{l}$; and 3) we modify the Laplacian matrix $\hat{\Lmat}^{l}$, such that $\hat{\Lmat}^{l+1}$ excludes the $\hat{k}$ edge.
    Otherwise,
the algorithm stops.
    The rationale behind the stop condition in (\ref{stopcondition}) is that the zero value is associated with the null hypothesis, $\mathcal{H}_0$ (of no disconnections), since the likelihood ratio satisfies $l(\yvec|\Lmat^{({0})})-\rho(\Lmat^{({0})})=0$.
 If the maximum of edge disconnections, $r_{\max}$, is known, then it can be used as an additional stopping condition.
 In addition, a restriction to a specific set of possible edge  disconnections,
 $\mathcal{C}^{(k)}$, $k=0,1,\ldots,K$, can be done by  adding a projection step at the end of Algorithm \ref{greedy_al}.
\begin{algorithm}[hbt]
\DontPrintSemicolon
   \KwInput{\begin{itemize} 
   \item Sample covariance matrix, $\Smat_{\yvec}$ 
   \item  Initial Laplacian matrix, $\Lmat^{(0)}$, and its
 edge set, $\mathcal{E}^{(0)}$
   \item Signal and noise variances, $\sigma_{\xvec}^2$ and $\sigma_{\wvec}^2$
   \item Graph filter, $h(\cdot)$
   \item Optional: Sparsity level, $r_{\max}$.
   \end{itemize}
  }
  \KwOutput{Estimated edge disconnections set, $\hat{\mathcal{C}}$.}
   Initialize $\hat{\mathcal{C}}^0=\emptyset$, $\hat{\mathcal{E}}^{0}=\mathcal{E}^{(0)}$, $\hat{\Lmat}^{0}=\mathcal{\Lmat}^{(0)}$, and $l=0$.
   
   Find the maximal edge, $\hat{k}\in \hat{\mathcal{E}}^{l}$, by \eqref{greedy_likelihood}. \label{2_1}
     \\
            \If{ $l(\yvec|\hat{\Lmat}^{l}-\Emat^{(\hat{k})})-\rho(\hat{\Lmat}^{l}-\Emat^{(\hat{k})})\:>\:0$}
            {Update $\hat{\mathcal{C}}^{l+1}= \hat{\mathcal{C}}^l\cup \{\hat{k}\}$,  $\hat{\mathcal{E}}^{l+1}=\hat{\mathcal{E}}^{l}\setminus\{\hat{k}\}$, 
             $\hat{\Lmat}^{l+1}=\hat{\Lmat}^{l}-\Emat^{(\hat{k})}$, and $l\leftarrow l+1$.\\
            \If{$|\hat{\mathcal{C}}^l| = r_{\max}$
            }
            {\textbf{Return:} $\hat{\mathcal{C}}^l$.}
             \textbf{Repeat} to step \ref{2_1}.
            }
     \textbf{Return:} $\hat{\mathcal{C}}^l$.
        \caption{Greedy  identification}\label{greedy_al}
\end{algorithm}
\vspace{-0.25cm}
\subsection{Greedy approach with a neighboring strategy}\label{Neighboring Strategy}
The computational complexity of the greedy algorithm in Algorithm \ref{greedy_al} may still be too high for large networks.
This is due to the fact that, according to
 (\ref{LRT11}) and 
(\ref{rho p}),
 evaluating the r.h.s. on
(\ref{greedy_likelihood}) at each iteration,    requires computing  around $|\mathcal{E}^{(0)}|$ $N$-dimensional inverse matrices. 
In this section 
 we propose additional simplifications by computing the local statistics and searching only in the neighborhood of the suspected edges.
The neighboring strategy is inspired by the local property of the GMRF filter in Proposition \ref{pro_GMRF},
which states that we only need to consider the vertices  in  $\mathcal{S}^{(k)}$. Here, we  use  the vertices in the $\beta$-neighborhood of the suspicious edges for  general   smooth graph filters that are expected  to have similar values at neighboring vertices.
The tunable parameter $\beta$ is the number
of neighbors taken into consideration,  provides a trade-off between the identification accuracy  and the  computation cost.

The neighboring strategy comes to ease the greedy algorithm's complexity by: 1) calculating  the $\beta$-local ML decision rule for a single edge, where for a given candidate edge, $(i,j)\in\hat{\mathcal{E}}^{l}$, we calculate the likelihood ratio of the  measurements in the  set $\mathcal{N}((i,j),\beta)\define\mathcal{N}(i,\beta)\bigcup\mathcal{N}(j,\beta)$, where  $\mathcal{N}(i,\beta)$ is 
the set of vertices connected to vertex $i$ by a path of at most $\beta$  edges; 
 and 2) building the new set $\hat{\mathcal{E}}^{l+1}$ of the suspicious edges for the $l+1$th iteration, where for sparse graphs, $\hat{\mathcal{E}}^{l+1}$ is significantly smaller than $\mathcal{E}^{(0)}$, which is required for searching over all the edges in the graph.
This two-step approach is as follows.
\subsubsection{Calculating the $\beta$-local ML decision rule}
In order to compute the likelihood locally, 
the greedy iteration from (\ref{greedy_likelihood}) is replaced by the local ML decision rule that we define by 
\begin{align*}\label{MML2}
   \hat{k}=\argmax_{k=(i,j)\in \hat{\mathcal{E}}^{l}}&\Phi_1(\yvec,\hat{\Lmat}^{l},\Emat^{(i,j)},\mathcal{N}((i,j),\beta))\\
   &-\Phi_2(\hat{\Lmat}^{l},\Emat^{(i,j)},\mathcal{N}((i,j),\beta)) , \numberthis
  \end{align*}
  where
  \begin{align*}\label{phi_1}
    \Phi_1(\yvec,\Lmat,\Emat, &\mathcal{S})  \triangleq  
    \text{Tr}\bigg(\big(\sigma^2_{\xvec}[h^2({\Lmat})]_{\mathcal{S}}+\sigma^2_{\wvec}\Imat\big)^{\dagger}[\Smat_{\yvec}]_{\mathcal{S}}\bigg)
        \\&
    -\text{Tr}\bigg(\big(\sigma^2_{\xvec}[h^2({\Lmat}-\Emat)]_{\mathcal{S}}+\sigma^2_{\wvec}\Imat\big)^{\dagger}[\Smat_{\yvec}]_{\mathcal{S}}\bigg)\numberthis
  \end{align*}
  and 
  \begin{align*}\label{phi_2}
     &\Phi_2(\Emat,{\Lmat},\mathcal{S}) \triangleq 
    \log\bigg(\frac{|\sigma_{\xvec}^2[h^2({\Lmat}-\Emat)]_{\mathcal{S}}+\sigma^2_{\wvec}\Imat|_+}{|\sigma_{\xvec}^2[h^2({\Lmat})]_{\mathcal{S}}+\sigma^2_{\wvec}\Imat|_+}\bigg).\numberthis
  \end{align*}
  The expressions in (\ref{phi_1}) and (\ref{phi_2}) are the local equivalent to the expressions in (\ref{LRT11}) and (\ref{rho p}), respectively.  For the special case where $\mathcal{N}((i,j),\beta)=\mathcal{V}$, which is obtained by taking a  maximal value of $\beta$,   \eqref{MML2} is reduced to 
  (\ref{greedy_likelihood}) from
 the greedy approach in Algorithm  \ref{greedy_al}.

\subsubsection{Building a suspicious edge set} In order to build a new subset of suspicious edges, we initialize this set to $\hat{\mathcal{E}}^{0}=\mathcal{E}^{(0)}$. 
Then,
  in the $l$th iteration, we determine the search edge set for the $l+1$ iteration by
            \begin{align*}\label{NN}
          \hat{\mathcal{E}}^{l+1}=
          \big\{{\mathcal{E}}_+^{l}\bigcup {\mathcal{E}}_\beta^{l}\big\}
          \setminus \hat{\mathcal{C}}^l,
    \numberthis
\end{align*}
where 
\begin{align*}\label{NNp}
          {\mathcal{E}}_+^{l}\define\bigg\{(i,j)\in& \hat{\mathcal{E}}^{l}|
         \Phi_1(\yvec,\hat{\Lmat}^{l},\Emat^{(i,j)},\mathcal{N}((i,j),\beta))\\
          &-\Phi_2(\hat{\Lmat}^{l},\Emat^{(i,j)},\mathcal{N}((i,j),\beta))>0\bigg\},
    \numberthis
\end{align*}
in which $\Phi_1(\cdot)$ and $\Phi_2(\cdot)$ are defined in (\ref{phi_1}) and (\ref{phi_2}), respectively, 
and
\begin{align*}\label{PPbeta}
 {\mathcal{E}}_\beta^{l}\define\bigg\{(i,j)\in \mathcal{E}^{(0)}|\exists(u,v)\in\big\{ \big\{\mathcal{P}(i,\beta)\cup\mathcal{P}(j,\beta)\big\}\cap
         {\mathcal{E}}_+^{l}\big\}
           \bigg\},
               \numberthis
\end{align*}
 in which  
 $\mathcal{P}(i,\beta)$ is the set of edges in the shortest paths between the vertix $i$ and the vertices in $\mathcal{N}(i,\beta)$.
 That is, 
 we include in  the search set  in (\ref{NN})
only
edges: a) with a non-negative $\beta$-local likelihood ratio, which is the term on the r.h.s. in (\ref{MML2}), i.e. edges in  ${\mathcal{E}}_+^{l}$, defined in \eqref{NNp};
and b) edges that are in the $\beta$-neighborhood of such edges as in a), i.e. edges in 
 ${\mathcal{E}}_\beta^{l}$, defined in \eqref{PPbeta}.
 The rationale behind this set  is that the zero value is associated with the null hypothesis (of no disconnections). Thus, suspicious edges should have non-negative values or at least  be in the local neighborhood of such edges. 

The two-step greedy algorithm with neighboring strategy is summarized  in Algorithm \ref{greedy_al_nig}.
\begin{algorithm}[hbt]
\DontPrintSemicolon
   \KwInput{
     \begin{itemize} 
   \item Sample covariance matrix, $\Smat_{\yvec}$ 
   \item  Initial Laplacian matrix, $\Lmat^{(0)}$, and its
 edge set, $\mathcal{E}^{(0)}$
   \item Signal and noise variances, $\sigma_{\xvec}^2$ and $\sigma_{\wvec}^2$
   \item Graph filter, $h(\cdot)$
   \item   Neighboring order, $\beta$
   \item Optional: Sparsity level, $r_{\max}$
   \end{itemize}
  }
  \KwOutput{The estimated edge disconnections set $\hat{\mathcal{C}}$.}
   Initialize $\hat{\mathcal{C}}^0=\emptyset$,  $\hat{\mathcal{E}}^{0}=\mathcal{E}^{(0)}$,  $\hat{\Lmat}^{0}=\Lmat^{(0)}$ and $l=0$.\\
    Find the maximal edge, $\hat{k}\in \hat{\mathcal{E}}^{l}$, by \eqref{MML2}.
    \label{2_2}\\
            \If{ $\Phi_1(\yvec,\hat{\Lmat}^{l},\Emat^{\hat{k}},\mathcal{N}(\hat{k},\beta)) -\Phi_2(\hat{\Lmat}^{l},\Emat^{\hat{k}},\mathcal{N}(\hat{k},\beta))>0$}
            {Update $\hat{\mathcal{C}}^{l+1}= \hat{\mathcal{C}}^{l}\cup \{\hat{k}\}$ and $\hat{\Lmat}^{l+1}=\hat{\Lmat}^{l}-\Emat^{\hat{k}}$.\\ Build the  edge set $\hat{\mathcal{E}}^{l+1}$ by (\ref{NN}), and $l\leftarrow l+1$. \\
            \If{$|\hat{\mathcal{C}}^l| = r_{\max}$
            }
            {\textbf{Return:} $\hat{\mathcal{C}}^l$.}
             \textbf{Repeat} to step \ref{2_2}.
            }
    { \textbf{Return:} $\hat{\mathcal{C}}^l$.}
        \caption{Greedy identification with a neighboring Strategy}
        \label{greedy_al_nig}
\end{algorithm}
\vspace{-0.25cm}
\subsection{Remarks}\label{Advantages}
\subsubsection{Computational complexity}
\label{comp_subsec}
The computational complexity of the proposed greedy algorithms is significantly lower than those of the original ML decision rule due to the following reasons. 
First, in the general worst-case scenario, where any of the edges may be disconnected,  the ML decision rule from (\ref{MML min1}) requires $K$ calculations of the likelihood ratio, where $K$ is a combinatorial term defined in (\ref{KK}). In contrast, Algorithms \ref{greedy_al} and  \ref{greedy_al_nig} are based on a search approach described in (\ref{greedy_likelihood}) and (\ref{MML2}), respectively, that are performed over $r_{\max}\times |\mathcal{E}^{(0)}|$ (or less) possibilities, which is significantly smaller than $K$ for large networks. 
 Second, if we  assume that the considered graphs are
sparse  with a small degree,  such that $|\mathcal{E}^{(0)}|\ll \frac{N(N-1)}{2}$, then, for  small values of $\beta$, most of the edges have small sets of their $\beta$-local neighborhood, that are used in Algorithm \ref{greedy_al_nig}.  In particular,   the computational complexity of inverting matrices in  Algorithm \ref{greedy_al_nig} is $\mathcal{O}(|\mathcal{N}((i,j),\beta)|^3)$, since we have smaller matrices  in (\ref{phi_1}) and (\ref{phi_2}), where the size depends on the dimension of the $\beta$- neighborhood of the specific edge $(i,j)$.
This is in contrast with the computational complexity of the ML decision rule and Algorithm \ref{greedy_al} that require computing the inverse of the $N$-dimensional   matrices in (\ref{MML min1}) and in (\ref{greedy_likelihood}), which  is $\mathcal{O}(N^3)$. 
Third, 
in Algorithm \ref{greedy_al_nig},
the size of the searching edge set    in \eqref{NN} tends to be smaller than the size of   searching edge set
of Algorithm \ref{greedy_al}, $ |\mathcal{E}^{(0)}|$,
as long as the graph is sparse.
Finally, both the ML decision rule and the greedy approaches require EVD calculation. Recent works propose low-complexity  methods to reduce the complexity of this task (see, e.g. \cite{SVD}).
{\color{black}Finally, the computational complexity of the proposed methods is summarized in Table \ref{complexity}.   }
 \begin{table}[hbt]
     \centering
  \begin{tabular}{ |m{1.3cm}|m{2cm}|m{2cm}|m{2cm}|  }
 \hline
 &{\color{black}\# Likelihood ratio calculation }&{\color{black}Matrices inversion}&{\color{black}Searching edge set size}\\
 \hline
 {\color{black} ML decision rule}&{\color{black} $\sum\limits_{r=1}^{r_{\max}}{|\mathcal{E}^{(0)}| \choose r}$}    &  {\color{black}$\mathcal{O}(N^3)$} &   {\color{black}$\sum\limits_{r=1}^{r_{\max}}{|\mathcal{E}^{(0)}| \choose r}$} \\
 \hline
  {\color{black} Greedy approach}&  {\color{black}$r_{\max}\times |\mathcal{E}^{(0)}|$ } & {\color{black} $\mathcal{O}(N^3)$ }  &  {\color{black}$|\mathcal{E}^{(0)}|$ }\\\hline
   {\color{black} Greedy approach with neighboring strategy}&{\color{black} $r_{\max}\times |\mathcal{E}^{(0)}|$} &  {\color{black} $\mathcal{O}(|\mathcal{N}((i,j),\beta)|^3)$} &   {\color{black}$|\hat{\mathcal{E}}^{l}||$ }\\
 \hline
\end{tabular}
\caption{{\color{black}The computational complexity of the proposed methods. }}
 \label{complexity}
\end{table}

\subsubsection{Local LRT for binary hypothesis testing}\label{LLRT_section}
Similar to Algorithm \ref{greedy_al_nig}, we can implement a low-complexity version of the LRT in (\ref{LRT1})-(\ref{LRT11}) by applying a neighboring strategy. In this case, the low-complexity local LRT for testing the disconnections in of the edges in ${\mathcal{C}}^{(k)}$ is based only on measurements in the  set ${\textstyle \bigcup_{(i,j)\in\mathcal{C}^{(k)}}}\mathcal{N}((i,j),\beta)$, instead of using the full-graph measurements. That is,
$ l(\yvec|\Lmat^{(k)})$ 
from \eqref{LRT1}-\eqref{LRT11} 
is replaced 
 by 
    \begin{align*}\label{LLRT}
        \Phi_1\bigg(\yvec,\Lmat^{(0)},\Emat^{(k)},{\textstyle \bigcup_{(i,j)\in\mathcal{C}^{(k)}}}\mathcal{N}((i,j),\beta)\bigg)\lessgtr^{\mathcal{H}_0}_{\mathcal{H}_1}\gamma' ,\numberthis
    \end{align*}
    where $\Phi_1(\cdot)$ is defined in (\ref{phi_1}) and $\Emat^{(k)}$ is associated with the set $\mathcal{C}^{(k)}$ and is defined in (\ref{multierror1}).

\section{Simulations}\label{Simulations}
In this section, we evaluate the performance of the LRT 
and the greedy approaches from Sections \ref{Detection of Topology Change} and \ref{Proposed Solutions},  and  compare them with the performance  of state-of-the-art methods.
In Subsection \ref{Experimental Settings}, we describe the general experimental setting {\textcolor{black}{for the synthetic data}} and methods for comparison. In Subsections \ref{detection} and \ref{identification}, we demonstrate  the  performance of the detection and identification approaches, respectively. {\color{black}Finally, we present simulations concerned with  the identification of
outages in  power systems  in Subsection \ref{power}.}

\subsection{Experimental settings and methods for comparison}\label{Experimental Settings}
In  the simulations in Subsections \ref{detection} and \ref{identification}, 
the data   was generated according to the measurement model from Subsection \ref{Measurements Model}, where
 the  initial graph, $\mathcal{G}^{(0)} = (\mathcal{V}, \mathcal{E}^{(0)},\Wmat^{(0)} )$, were  generated by using the Watts-Strogatz small-world graph model \cite{Watts},   with  $N=50$ vertices,  mean degree of $K=2$, and  $|\mathcal{E}^{(0)}|=100$. {\textcolor{black}{Similar results were obtained for other random graphs, such as graphs generated by the stochastic block model. These results were omitted from this paper due to space limitations.}}
The elements of
the adjacency matrix, $\Wmat^{(0)}$, are independent  uniform distributed weights in the range $[0.1,5]$. The graph after the change is obtained by removing an arbitrarily chosen set of $r$ edges from  $\mathcal{E}^{(0)}$.
The output graph signals  were generated by
implementing \eqref{model},  where $\sigma_{\xvec}^2=1$. 
The graph filters from Table \ref{FT} were tested with  parameters $\alpha=0.5$ and $\tau=0.2$.
At least $1,000$ Monte-Carlo simulations were conducted to evaluate the performance in each scenario.

The methods for comparison are as follows:
\subsubsection{Naive smoothness detector}  The naive smoothness detector measures the average smoothness of the output graph signal, $\yvec$, w.r.t. the initial Laplacian, $\Lmat^{(0)}$:
\begin{align*}\label{smooth naive}
    \hspace{-0.25cm}  \frac{1}{M}\sum_{m=1}^M Q_{\Lmat^{(0)}}(\yvec[m])=\frac{1}{M}\sum_{m=1}^M \yvec^{T}[m]\Lmat^{(0)} \yvec[m]\numberthis \lessgtr^{\mathcal{H}_0}_{\mathcal{H}_1}\gamma,
 \end{align*}
 where $\gamma$ is a chosen threshold. 
The underlying assumption behind this detector is that if this quantity obtains a small value, then, it is more likely that  $\yvec$ has been generated from $\mathcal{G}^{(0)}$; otherwise,  $\yvec$ is associated with a different topology.
{\textcolor{black}{In contrast with the  LRT    for the  GMRF  filter  in (\ref{smooth1}), the naive smoothness detector in \eqref{smooth naive} does not assume that the topology after the change, represented by $\Lmat^{(k)}$, is known. }}
\subsubsection{Matched subspace  detectors (MSDs) \cite{Matched,Elvin}} 
The MSDs  compare the energies of a graph signal $\yvec$ at high graph frequencies of the graph spectral domain 
 under each hypothesis.
We implement here
1) the 
simple-MSD (SMSD):
\beqna\label{SMSD}
    &\frac{1}{M}\sum_{m=1}^M \left( \norm{\tilde{
    \yvec}^{
    \Lmat^{(0)}}_{\mathcal{V}\setminus {\mathcal{S}}_B}[m]}^2_2-\norm{\tilde{
    \yvec}^{
    \Lmat^{(k)}}_{\mathcal{V}\setminus {\mathcal{S}}_B}[m]}^2_2\right)\lessgtr^{\mathcal{H}_0}_{\mathcal{H}_1} \gamma,
\eeqna
where $\mathcal{S}_B=
1,\ldots,B$, and ${\mathcal{H}_0},{\mathcal{H}_1}$ are the hypotheses in \eqref{hypothesis testing}; 
and 2) the blind simple-MSD (BMSD): \begin{equation}\label{BMSD}
   \frac{1}{M}\sum_{m=1}^M\norm{\tilde{
    \yvec}^{
    \Lmat^{(0)}}_{\mathcal{V}\setminus {\mathcal{S}}_B}[m]}^2_2\lessgtr^{\mathcal{H}_0}_{\mathcal{H}_1}\gamma.
\end{equation}
 We determine the parameter to be $B=\lceil\frac{N}{4}\rceil$. 
The difference between the SMSD and the BMSD is that the SMSD does 
assume that the topology after the change is known.
\subsubsection{Combinatorial graph Laplacian (CGL) method \cite{Eduardo,Egilmez}} \label{CGL_section}
The CGL  is a  Laplacian learning block-coordinate descent  method, which is based  on the Laplacian structure constraints.
{\textcolor{black}{The CGL method was shown to be useful in cases of up to 25\%  mismatch in the connectivity of the graph (see Fig. 6 in \cite{Eduardo}). In addition, the model behind the CGL method in \cite{Eduardo,Egilmez} results in the same  covariance matrix  as the covariance matrix of  the considered model in (\ref{Sigma}). Therefore, the CGL method is appropriate for comparison in the case of problems dealing with  identification of edge disconnections in networks based on graph filter outputs.
}}
It is based on
the connectivity matrix of $\mathcal{G}$,  $\Amat$, which has the following $(i,j)$ element
\be
A_{i,j}\triangleq\left\{ \begin{array}{cc}
    1 &  {\text{if }} W_{i,j}\neq 0 \\
    0 & {\text{otherwise }}
\end{array}\right.,
\ee
 where $\Wmat$ is the adjacency matrix of $\mathcal{G}$. 
 In the following simulations, 
edge disconnection identification by the CGL method is implemented by the following four-step approach: 
\begin{enumerate}[(a)]
    \item  A prefiltering operation is performed, as described in \cite{Egilmez}, by removing the noise variance and using the inverse graph filter, $h^{-1}(\cdot)$, on the sample covariance matrix:
    \begin{align*}\label{prefiltered}
        \Smat^{\text{pf}}_{\yvec}\triangleq h^{-1}\bigg(\sqrt{\frac{1}{\sigma_\xvec^2}(\Smat_{\yvec}-\sigma_\wvec^2\Imat)}\bigg).\numberthis
    \end{align*}
    \item The CGL method is implemented  by using  the code  in \cite{Eduardo}
    with the inputs: 1) the connectivity matrix of the known topology  $\mathcal{G}^{(0)}$; and 2) the prefiltered sample covariance matrix, $\Smat^{\text{pf}}_{\yvec}$ from (\ref{prefiltered}),  and under the constrained set
 \begin{align*}
 \label{Lc}
     \mathcal{L}_{c}(\Amat)=\bigg\{ \Lmat\bigg|&\Lmat	\succeq \Omat,\:\Lmat\onevec=\zerovec,L_{i,j}\leq0 \:\: {\text{for }}i\neq j,\\
 & L_{i,j}=0 \:\: \text{if} \:\:
 A_{i,j}=0\bigg\}\numberthis{}.
 \end{align*}
    \item The off-diagonal elements of the estimator of $\Lmat$ from Step (b) are thresholded, such that elements that are larger than $-\epsilon$ are set to zero,  where we set $\epsilon=0.1$.
\item For any edge $(i,j)\in \mathcal{E}^{(0)}$, where the output of Step (c) satisfies $\hat{{{L}}}_{i,j}=0$ but $ L^{(0)}_{i,j}\neq 0$,  $(i,j)$ is declared as a disconnection. 
\end{enumerate}
{\color{black} For sparse graphs, the computational complexity of the CGL method is $\mathcal{O}(\Omega(d^3)+N^2)$, where $d=\max_{v\in\mathcal{V}}|\mathcal{N}(v,1)|$ is the maximum degree of the graph \cite{Egilmez}. }
\subsubsection{Constrained CGL (CCGL) method}\label{CCGL sec}
In order to have a fair comparison with the methods proposed in this paper, 
we also implement
the CCGL method, which is based on the CGL method after
 adding the information regarding the initial Laplacian matrix, $\Lmat^{(0)}$. This information has been integrated into the CGL method  by replacing \eqref{Lc} in Step (b) by the constraints set
\begin{align*}\label{CCGL}
\mathcal{L}_{cc}(\Amat)=\bigg\{ \Lmat\bigg|&\Lmat	\succeq \Omat,\:\Lmat\onevec=\zerovec,\:L^{(0)}_{i,j}\leq L_{i,j}\leq0 \:\: \text{if} \:\:
A_{i,j}=1,\\
 & L_{i,j}=0 \:\: \text{if} \:\:
 A_{i,j}=0 \bigg\}.\numberthis{}
\end{align*}
 In addition, in Step (c), we change the thresholding by setting an individual thresholding for each entry in $\hat{\Lmat}$, $\epsilon_{i,j}=\frac{1}{2}L^{(0)}_{i,j}$.
 {\color{black}\subsubsection{GGM-GLRT  \cite{GGM}}\label{GLRT_sec}
 The GGM-GLRT  method  for deciding if an edge $(i,j)\in\mathcal{E}$ is connected or disconnected is \cite{GGM}:
\begin{align*}    \label{GLRT_detector}
     &\left[\frac{|[\Smat_{\yvec}]_{\mathcal{N}((i,j),1)\setminus\{i,j\}}|_+|[\Smat_{\yvec}]_{\mathcal{N}((i,j),1)}|_+}{|[\Smat_{\yvec}]_{\mathcal{N}((i,j),1)\setminus\{i\}}|_+|[\Smat_{\yvec}]_{\mathcal{N}((i,j),1)\setminus\{j\}}|_+}\right]^{M}\lessgtr^{{\mathcal{H}}_0^{(i,j)}}_{{\mathcal{H}}_1^{(i,j)}}\gamma,\numberthis
\end{align*}
where  ${H}_1^{(i,j)}$ represents the hypothesis of a disconnection in edge $(i,j)$, and  ${H}_0^{(i,j)}$ represents  the hypothesis that  $(i,j)$ is a connected edge. It can be seen that  the GGM-GLRT detector on  \eqref{GLRT_detector} is only a function of the first-order neighbors of the tested edge. For the identification problem, we  implement the GGM-GLRT in (\ref{GLRT_detector}) on all the edges in the original graph, i.e.  $\forall (i,j)\in \mathcal{E}^{(0)}$, where the  threshold $\gamma$ is experimentally determined.
The computational complexity of the GGM-GLRT as an identification method is $\mathcal{O}(|\mathcal{E}^{(0)}|d^3)$, where  $d=\max_{v\in\mathcal{V}}|\mathcal{N}(v,1)|$ is the maximum degree of the graph.
 It should be noted that   the model with the GMRF filter in (\ref{GMRF filter}) is a special case of GGM with the Laplacian as the precision matrix. However, the GGM-GLRT method does not assume a  Laplacian-based structure of the precision matrix and was developed for the noiseless case. Thus, it requires an accurate estimation of the covariance matrix by a large number of measurements, in contrast with the proposed methods.
}

\subsection{Detecting a specific set of edge disconnections }\label{detection}
\begin{figure*}[hbt]
\centering
\subfloat{\includegraphics[width=0.3\textwidth]{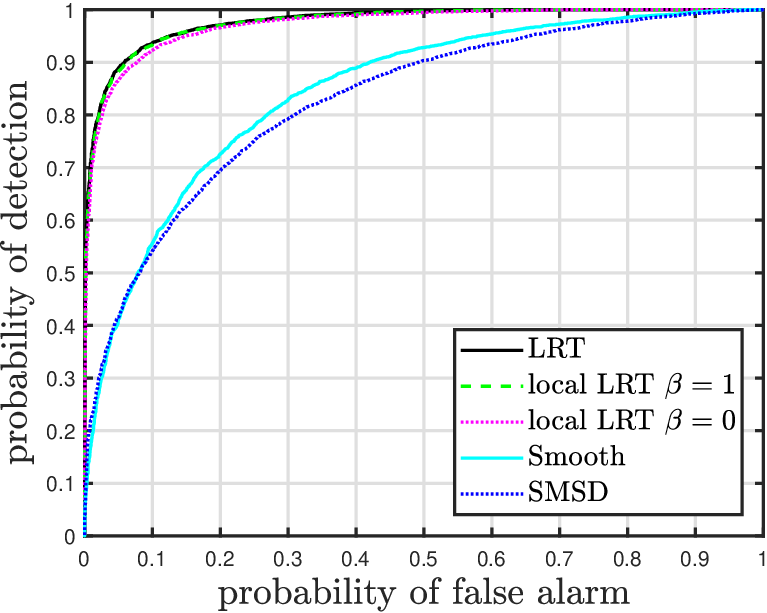}\label{fig:sub1}}
\subfloat{\includegraphics[width=0.3\textwidth]{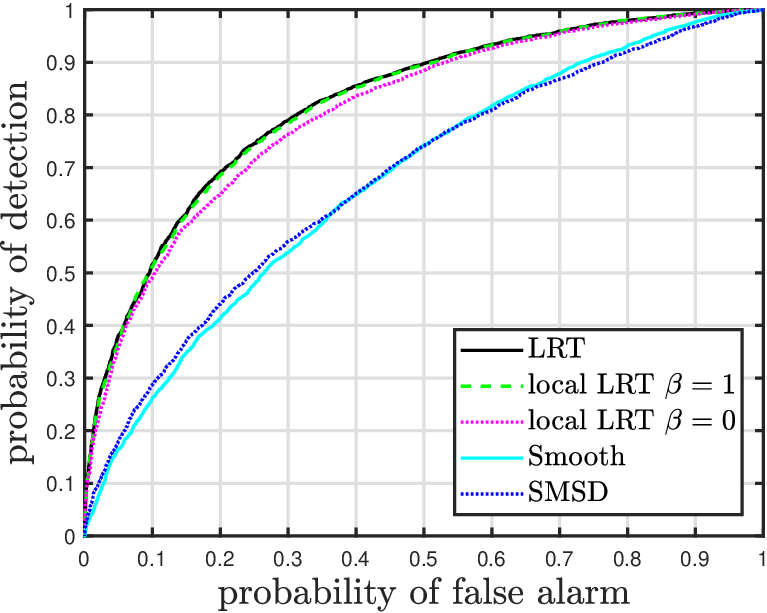}\label{fig:sub2}}
\subfloat{\includegraphics[width=0.3\textwidth]{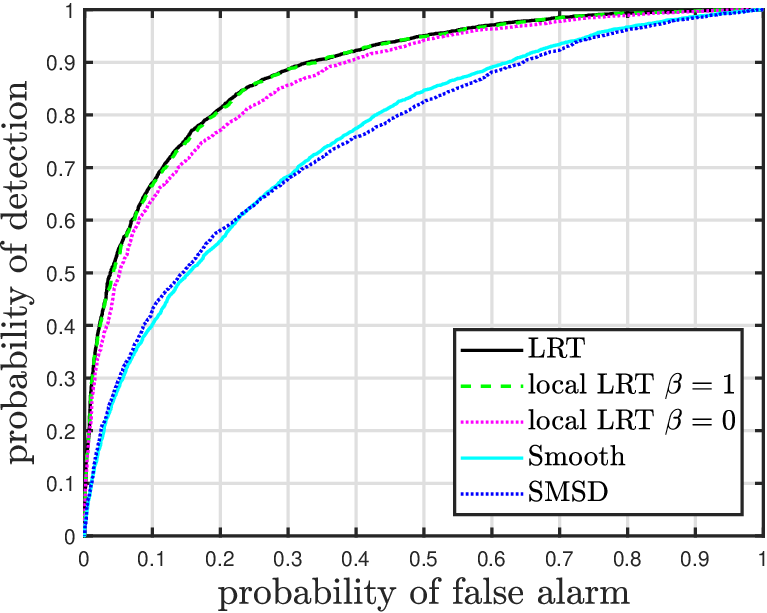}\label{fig:sub3}}
\caption{ROC curves of LRT, local LRT with $\beta=0,1$, naive smoothness detector, and SMSD, for: the GMRF filter (left), the regularized Laplacian filter (middle), and the heat diffusion filter (right), with  $\sigma_\wvec^2=2$, $M=100$ and $r=5$.}
\label{fig_LRT_comparation}
\end{figure*}

In this subsection, we evaluate the performance of tests that detect  a specific edge disconnection set for the graph filters from Table \ref{FT} with $r=5$ disconnections,  $M=100$ time samples, and noise variance $\sigma_{\wvec}^2=2$.
In
Fig. \ref{fig_LRT_comparation}, we present the Receiver Operating Characteristic (ROC) curves  of
the  detectors:  1) the LRT  in (\ref{LRT11}); 
2) the local LRT  in (\ref{LLRT})
 for  $\beta=0,1$; 
    3)  the naive smoothness detector in (\ref{smooth1}); and
    4) the SMSD detector in (\ref{SMSD}).
 It can be seen that the LRT and the local LRTs (with $\beta=0,1$) outperform the other detectors  for any probability of false alarm and for any tested smooth graph filter. 
In addition, the performance of the LRT and local LRT with $\beta=1$ is almost identical for any graph filter from Table \ref{FT}.  Thus, we can conclude that, in the considered cases, by taking into account only the first-order neighbors of the disconnected edges, we obtain the global performance that is based on measurements from the full graph. This property is due to the smoothness of the output graph signal and is a significant advantage when implementing in large networks.  

\subsection{Identifying edge disconnections}\label{identification}
In the following, we evaluate the performance of the proposed greedy algorithms from Subsection \ref{Proposed Solutions} in terms of 
detection and identification performance.
We assume the worst-case scenario, where any of the edges can be disconnected.
The results of the ML decision rule are not shown due to its computational complexity, which makes it impractical.

\begin{figure*}[hbt]
\centering
\subfloat{\includegraphics[width=0.3\textwidth]{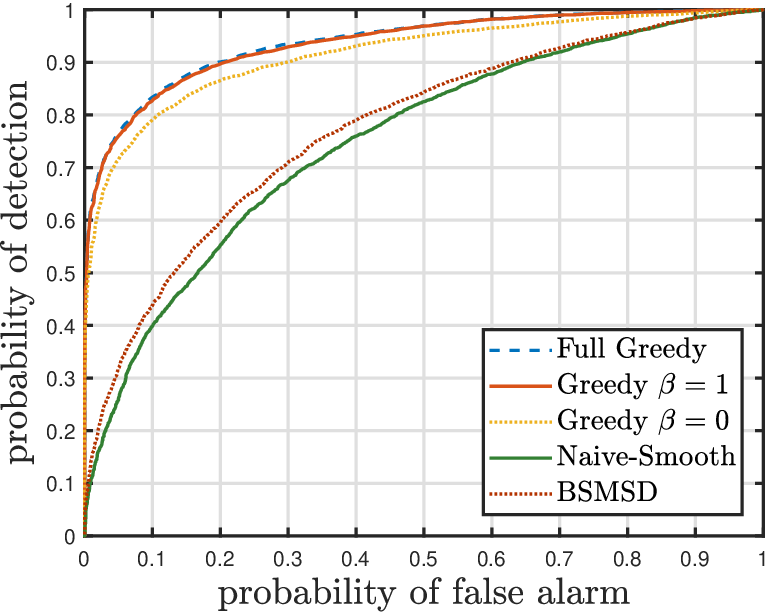}\label{fig:sub4}}
\subfloat{\includegraphics[width=0.3\textwidth]{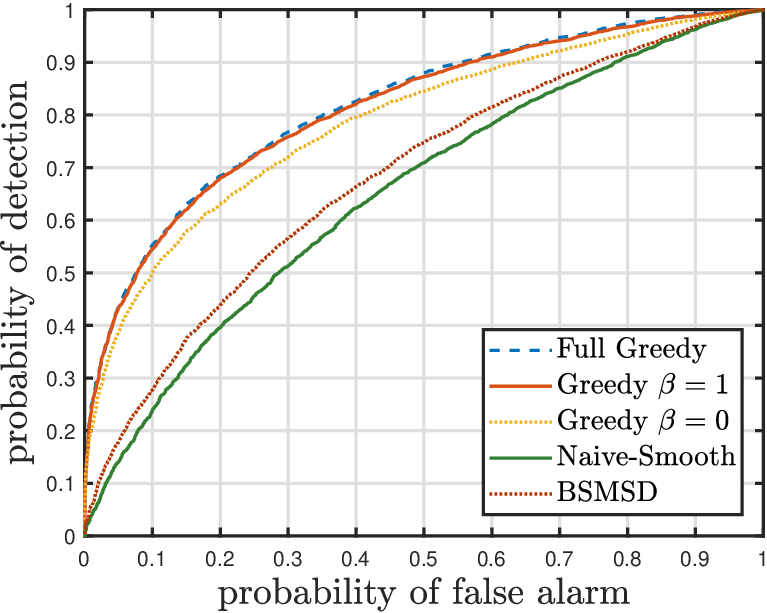}\label{fig:sub5}}
\subfloat{\includegraphics[width=0.3\textwidth]{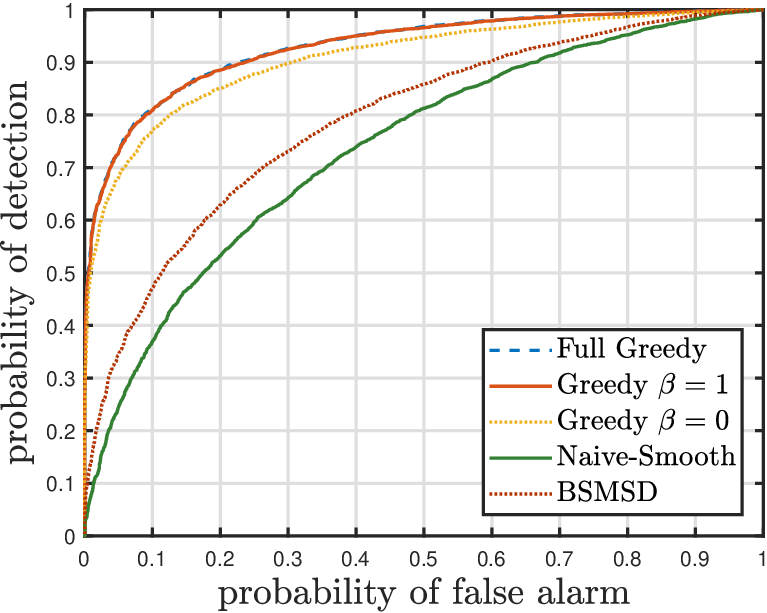}\label{fig:sub6}}
\caption{ROC curves of edge disconnections detector by estimating $
\hat{\Emat}$ with Algorithms \ref{greedy_al} and \ref{greedy_al_nig} with $\beta=0,1$, naive smoothness detector, and BMSD for: the GMRF filter (left), the regularized Laplacian filter (middle), and the heat diffusion filter (right), with noise variance $\sigma_\wvec^2=0.5$, $M=100$, and $r=5$.
}
\label{fig_UcGLRT_comparison}
\end{figure*}

In Fig.  \ref{fig_UcGLRT_comparison}, we evaluate  the performance of the detection question: is there any disconnection in the topology? That is, the null hypothesis is $\mathcal{H}_0$, while the alternative includes the union of all the other options,  $\mathcal{H}_1,\ldots,\mathcal{H}_K$. 
 According to the ML decision rule (\ref{MML min1}), we evaluate the performance of the detector
   \begin{align*}
   \label{59}
        l(\yvec|\Lmat^{(0)}-\hat{\Emat})-\rho(\Lmat^{(0)}-\hat{\Emat})\lessgtr^{\mathcal{H}_0}_{\mathcal{H}_1},\numberthis
    \end{align*}
 where $\hat{\Emat}$ is obtained from  the estimated edge disconnection, $\hat{\mathcal{C}}$, by  either the  greedy algorithm from Algorithm \ref{greedy_al} denoted  ``full" greedy or  by the greedy algorithm with the neighboring strategy in Algorithm \ref{greedy_al_nig} with  $\beta=0,1$. In each algorithm, we limit the 
 maximum number of possible edge disconnections in the network
 to $r_{\max}=10$.
 We compare these detectors with 
the naive smoothness detector  from (\ref{smooth naive})
and with the BMDS detector from (\ref{BMSD}).
Figure \ref{fig_UcGLRT_comparison} presents the ROC curves of these methods for the graph filters from Table \ref{FT} with $r=5$ disconnected edges,
a noise variance of $\sigma_{\wvec}^2=0.5$, and $M=100$ time samples.
 It can be seen that the edge disconnections detectors that were derived from Algorithms \ref{greedy_al} and \ref{greedy_al_nig} outperform the other detectors for any probability of false alarm and for any tested smooth graph filter.  In addition, similar to the results in Fig. \ref{fig_LRT_comparation},  the performance of the  greedy algorithm with neighboring strategy and $\beta=1$ is close to that  of the ``full" greedy algorithm. Thus, the first one should be preferred in practice since it has a lower computational complexity.


For evaluating the classification performance, we use the F-score measure \cite{Marina}:  \[\frac{\sum_{l=1}^L2tp_l}{\sum_{l=1}^L2tp_l+fn_l+fp_l},\] 
where $tp_l$, $fp_l$, $fn_l$ correspond to  the   true-positive, false-positive, and false-negative of the 
detection of disconnected edges in the $l$th simulation.
We compare the following topology identification methods:
1) the greedy approach in Algorithm \ref{greedy_al}; 2) the greedy approach with neighboring strategy in Algorithm \ref{greedy_al_nig} for $\beta=0,1$; 3) the CGL method;  4)
 the CCGL method; and {\color{black}5) the GGM-GLRT method}. The last {\color{black}three} methods are described at the end of Subsection \ref{Experimental Settings}.
 
 \begin{figure*}
    \centering
    \includegraphics[width=0.3\textwidth]{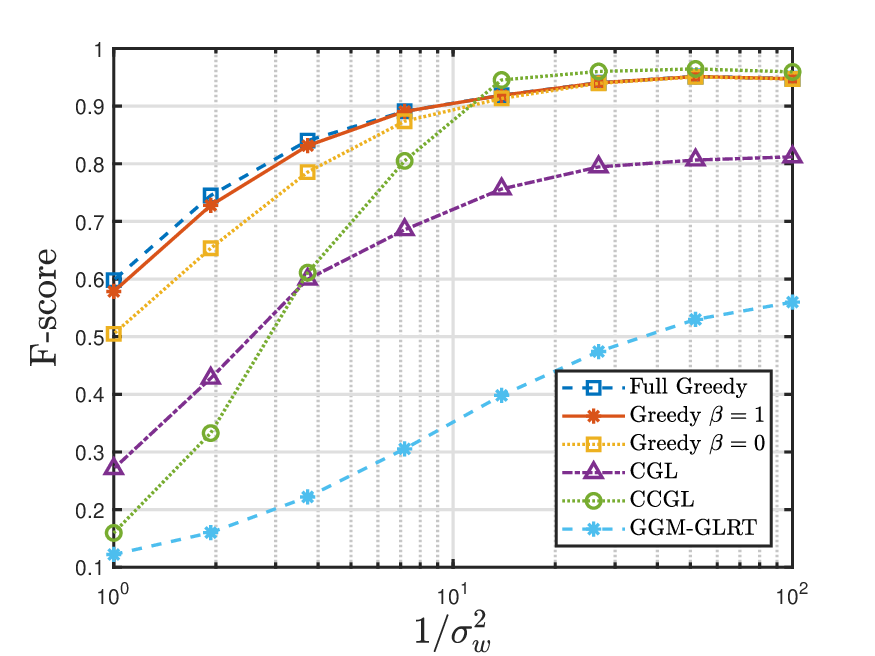}
    \includegraphics[width=0.3\textwidth]{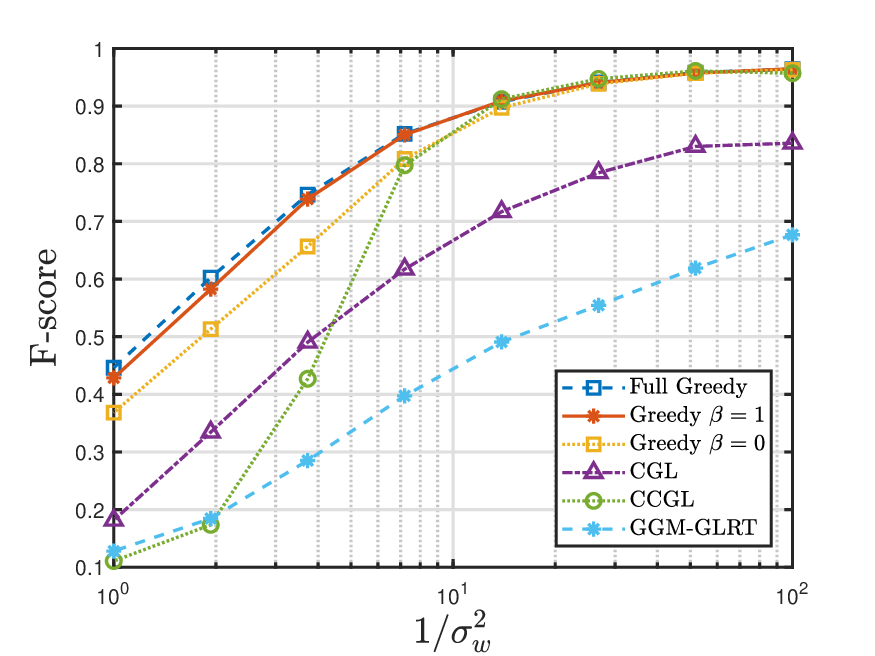}
    \includegraphics[width=0.3\textwidth]{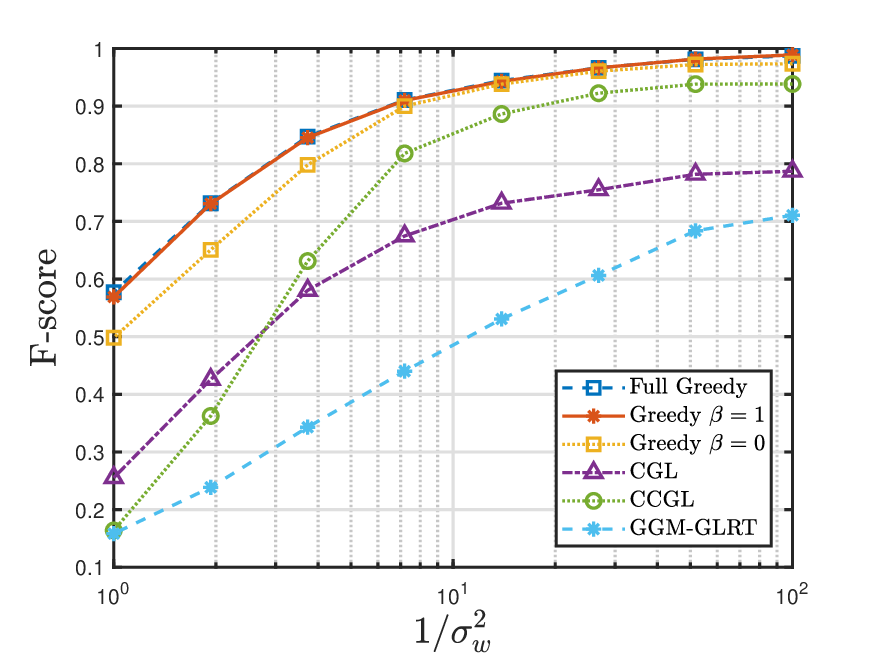}
    \caption{The F-score measure for each graph filter: the GMRF filter (left), the regularized Laplacian filter (middle), and the heat diffusion filter (right), versus SNR, $1/ \sigma_{\wvec}^2$, for the following methods: greedy approaches from Algorithm \ref{greedy_al} and Algorithm \ref{greedy_al_nig} for $\beta=0,1$, CGL method, the CCGL method, {\color{black} and the GGM-GLRT method} with $M=1,000$  and $r=5$. }
    \label{fig1}
\end{figure*}
\begin{figure*}
    \centering
    \includegraphics[width=0.3\textwidth]{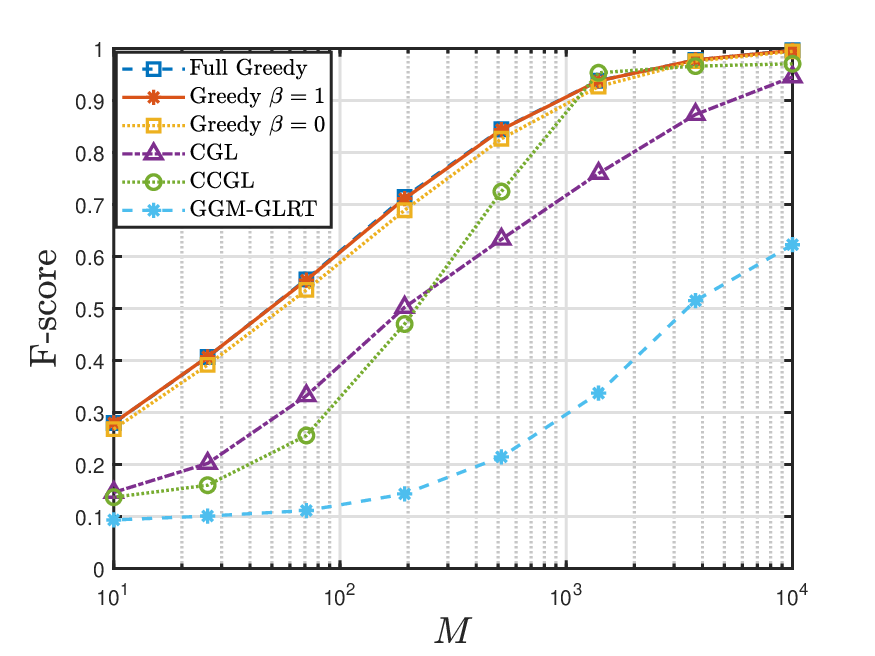}
    \includegraphics[width=0.3\textwidth]{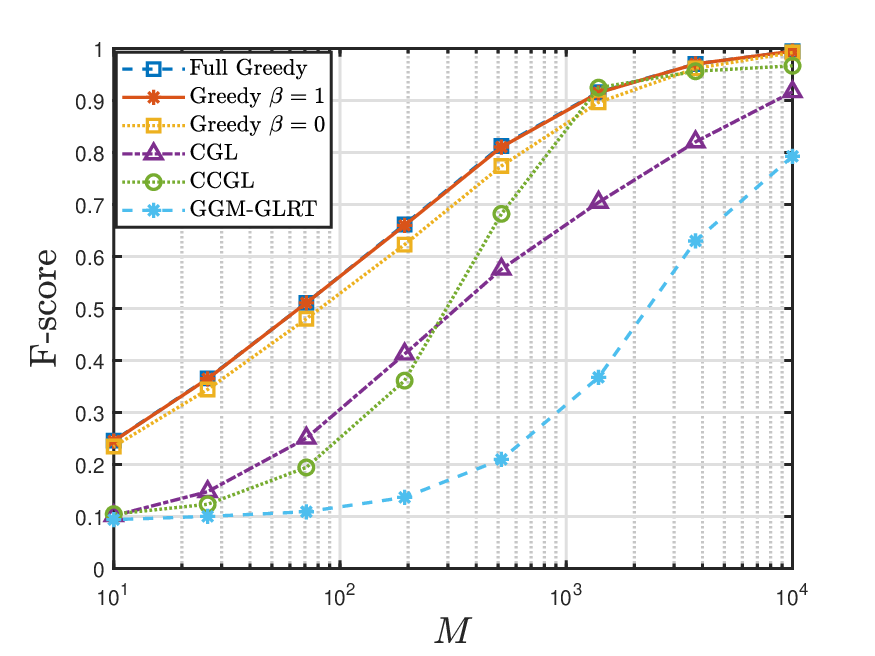}
    \includegraphics[width=0.3\textwidth]{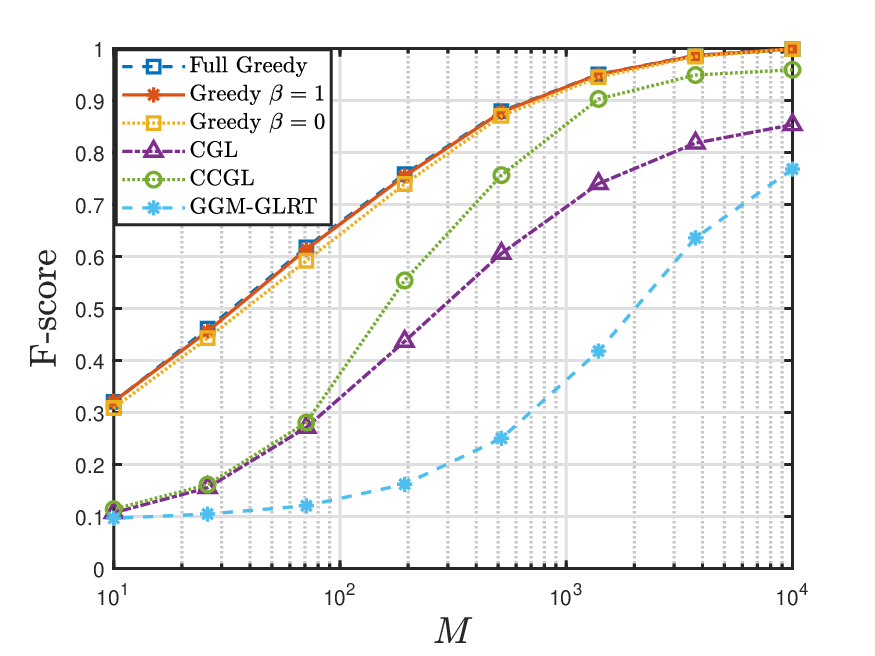}
    \caption{The F-score measure for each graph filter: the GMRF filter (left), the regularized Laplacian filter (middle), and the heat diffusion filter (right), versus number of measurements, $M$, for the following methods: greedy approaches from Algorithm 
    \ref{greedy_al} and Algorithm \ref{greedy_al_nig} for $\beta=0,1$, CGL method, CCGL method,  {\color{black} and the GGM-GLRT method} with $\sigma_w^2=0.1$  and $r=5$. }
    \label{fig2}
\end{figure*}
 
Figure \ref{fig1}  presents the F-score measure  of the different methods versus $\frac{1}{\sigma^2_{\wvec}}$ for  $M=1,000$ time samples and $r=5$.
Figure  \ref{fig2} presents the F-score measure  of the different methods versus  the number of measurements, $M$, for $\sigma^2_{\wvec}=0.1$ and $r=5$.
It can be seen that the F-score measure of all methods increases as the noise variance, $\sigma^2_{\wvec}$, decreases and/or where the number of time samples, $M$, increases.
In addition, 
 it can be seen that the proposed algorithms  outperform the other methods, where the  neighboring strategy with $\beta=1$ is preferred in terms of the trade-off between accuracy (it has almost the same performance as the ``full" greedy) and computational complexity (see in Subsection \ref{comp_subsec} and in the simulations below).  In addition, the proposed modification of the CGL method, the CCGL method, significantly improves the performance, compared with the original CGL method, for large $M$ and small $\sigma^2_{\wvec}$. Thus, considering the disconnection constraint in (\ref{CCGL}) and employing  appropriate thresholding  improves the CGL method from \cite{Eduardo,Egilmez} for the problem of edge disconnections identification. {\color{black}It can be seen that the GGM-GLRT method from \cite{GGM} is more sensitive to noise and to a small number of measurements than the other methods. This is because the model of the GGM-GLRT method does not consider the influence of the noise in the measurements, and is based on covariance matrix estimation, which requires a large number of measurements.}
  {\color{black}Finally, it can be seen that 
 the neighboring strategy, which was developed under the assumption of a local property of the GMRF filter (in Proposition \ref{pro_GMRF}), performs well for the regularized Laplacian filter and the heat diffusion filter in the middle and the right subfigures of Figs. \ref{fig_LRT_comparation}-\ref{fig2}. That is, the neighboring strategy is robust under the other tested smooth graph filters.}  

\begin{figure}[hbt]
\centering
\subfloat{\includegraphics[width=7cm]{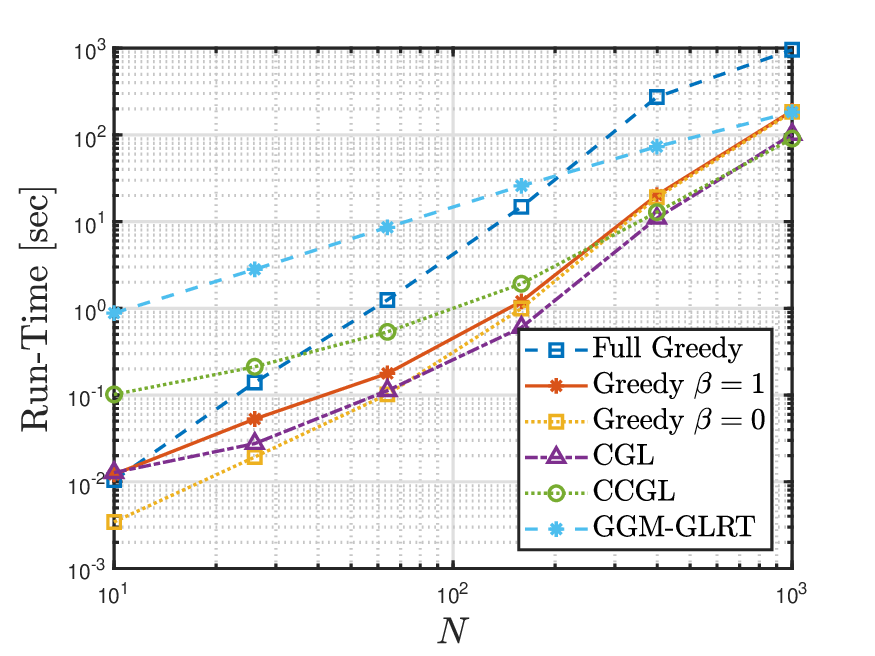}\label{fig:sub13}}
\caption{Run-time of the  methods:  greedy approaches from Algorithm \ref{greedy_al} and Algorithm \ref{greedy_al_nig} for $\beta=0,1$, CGL method, CCGL method, {\color{black} and the GGM-GLRT method} for $\sigma_\wvec^2=0.1$, $M=100,000$, $r=2$, and $L=100$.}
\label{fig3}
\end{figure}

In order to demonstrate the empirical complexity  of the identification methods for different network sizes, 
 the average computation time was evaluated by running these methods using Matlab on an Intel(R) Core(TM) i9-7900X CPU @ 3.30GHz. 
Figure \ref{fig3} shows the run-time of the different   methods versus the number of vertices in the graph, $N$,  for $\sigma_\xvec^2=0.1$, $M=100,000$, $r=2$, and $L=100$ Monte-Carlo simulations. It can be seen that the run-time of all methods increases as $N$ increases. For large values of $N$, the ``full" greedy approach from Algorithm \ref{greedy_al} has the largest run-time.  At the same time, the neighboring strategy efficiently reduces the run-time without a significant performance loss, as shown in the previous figures.  Moreover, for large $N$, the neighboring strategy obtains similar complexity to the CGL methods {\color{black}and the GGM-GLRT method}. The CCGL method was implemented by 'quadprog' solver in Matlab and its run-time can be reduced by implementing more efficient solutions, as proposed for the CGL in \cite{Eduardo,Egilmez}. 

{\color{black}\subsection{Identifying outages in power system dataset}\label{power}
In this subsection, we evaluate the performance of the proposed methods for  identifying line outages in electrical networks using  phasor angle measurements obtained by phasor measurement units (PMUs) \cite{Goldsmith,Giannakis,Wu,Tate_Overbye2008}.
A power system can be represented as an undirected weighted graph, $\mathcal{G}=(\mathcal{V}, \mathcal{E}, \Wmat)$, where the vertices in $\mathcal{V}$ denote the buses (generators or loads), and the edges in $\mathcal{E}$ denote transmission lines between these buses. The weighted adjacency matrix, $\Wmat$, is determined by the branch susceptances (as described, for example, in \cite{lital,Ariel2}).
We assume that we have PMUs in the considered system that acquire noisy  measurements of  the voltage phasors (amplitudes and phases)  at all buses. 
In addition, we assume that  the voltage amplitudes  equal one in a per unit (p.u.)  system, which is a common assumption in power systems \cite{Abur}.
Since the voltage phases are measured over the buses of the electrical network (vertices in the graph representation), they can be considered  as  graph signals. It has recently been shown in \cite{2021Anna,Drayer_2020,lital} that the voltages in power systems can be considered as smooth graph signals w.r.t. the associated  Laplacian matrix.
In order to obtain the voltage dataset, we  execute an optimal power flow solution obtained by  MATPOWER \cite{matpower} over  the power demand data embedded in Gaussian noise. 
 The simulations were implemented on the IEEE 118-bus test case, which is presented in Fig. \ref{fig:IEEE118} as a one-line diagram (left) and as a graphical representation
of this grid (right).
	\begin{figure*}[hbt]
    \centering
    \includegraphics[width=0.4\textwidth]{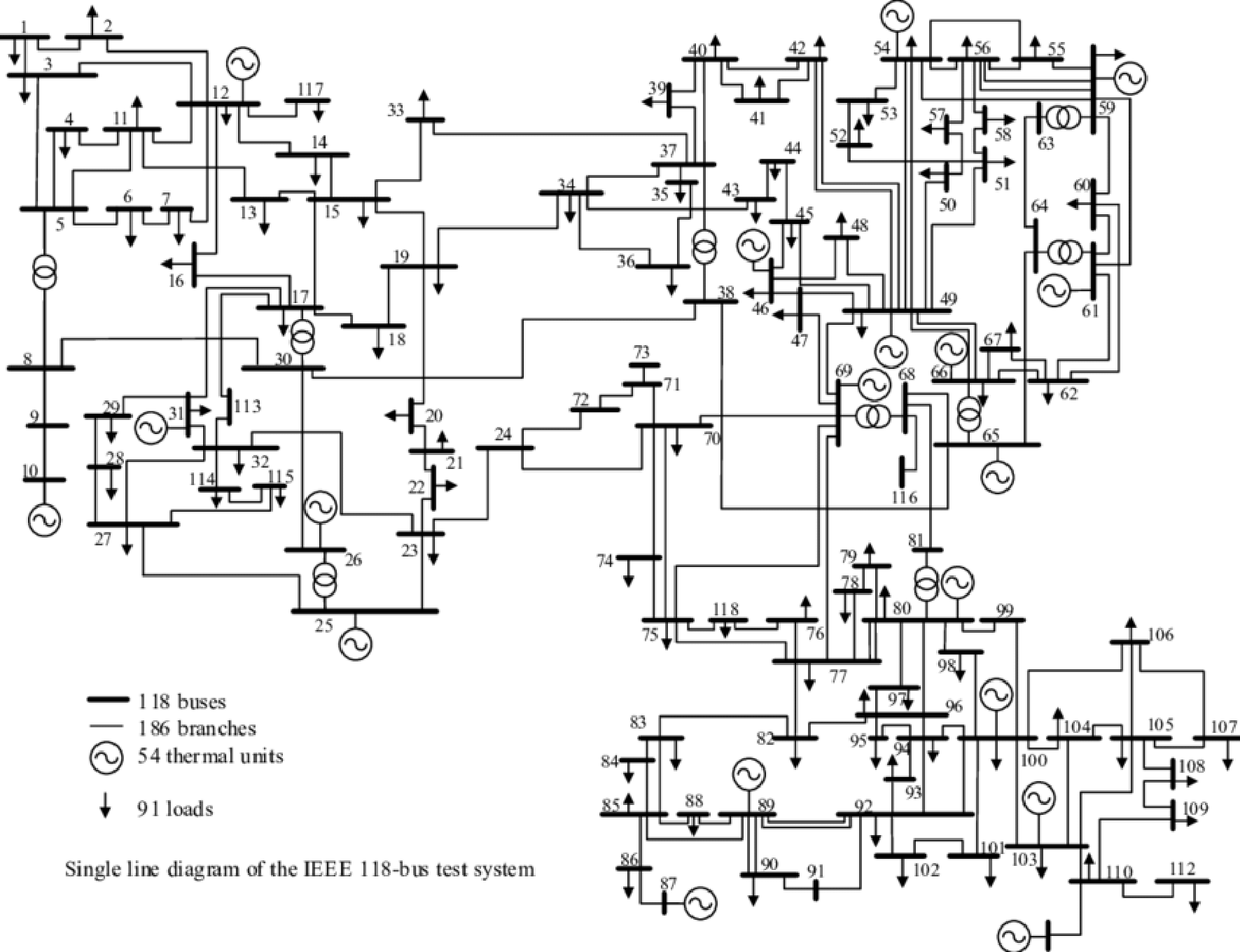}
    \includegraphics[width=0.45\textwidth]{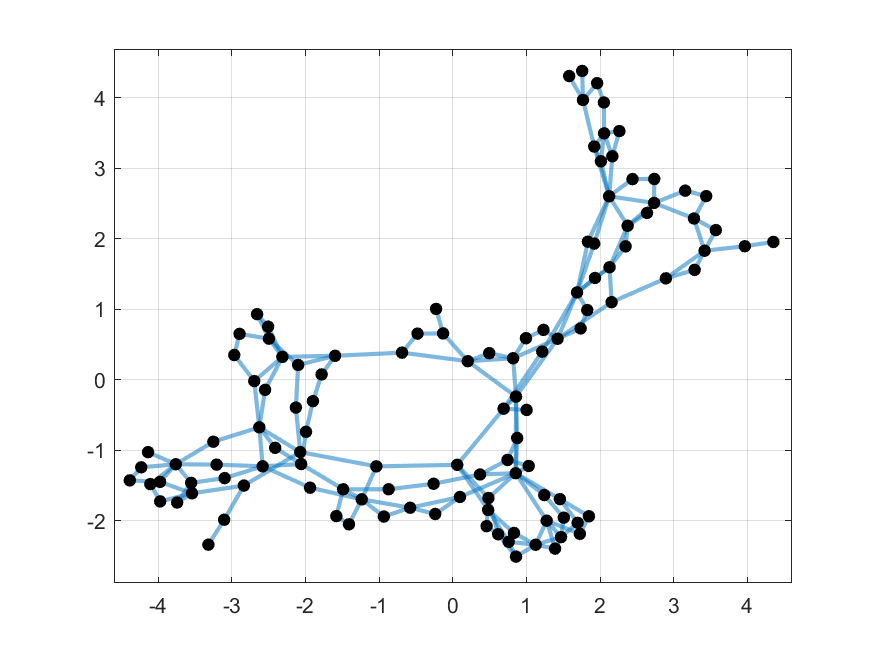}

    \caption{{\color{black}IEEE-118 bus test case \cite{matpower} (left)  and  its representation as a graph (right). }}
    \label{fig:IEEE118}
\end{figure*}

We tested four different outages at the transmission lines (disconnections of edges in the graph representation):  $\{(23, 24), (29, 31), (52, 53), (80,97)\}$. Then, we uniformly randomized one of the combinations of these outages such that $r\leq 4$ and $r_{\max}=10$.
 We implemented the proposed  greedy algorithms assuming a GMRF filter, which is a mismatch in the distribution of the signals. 
 
 Figure \ref{fig:ROCpower}  presents the ROC  
 of
the  detector in \eqref{59},
 where $\hat{\Emat}$ is obtained from  the  ``full" greedy algorithm or  by the greedy algorithm with the neighboring strategy in Algorithm \ref{greedy_al_nig} with  $\beta=0,1$,  for
a noise variance of $\sigma_{\wvec}^2=0.2$, 
and $M=1,000$ time samples;
 results from
the naive smoothness detector  from (\ref{smooth naive})
and with the BMDS detector from (\ref{BMSD}) are also shown.
 It can be seen that the  ``full" greedy algorithm outperforms the other detectors for any probability of false alarm. For the  greedy algorithm with neighboring strategy, the probability of detection increases as $\beta$ increases.
 The naive smoothness detector  
and the BMDS detector have a higher probability of detection than the $\beta=0,1$ greedy method for large probabilities of false alarm, which cannot be tolerated in practical systems.
}
	\begin{figure}[hbt]
    \centering
    \includegraphics[width=7cm]{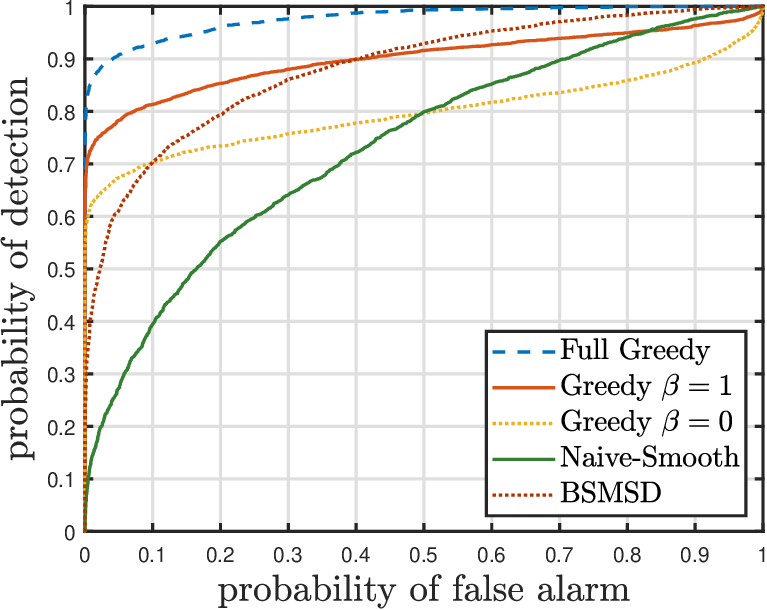}
    \caption{{\color{black}Identifying outages in power system: ROC curves of edge disconnections detector by estimating $
\hat{\Emat}$ with Algorithms \ref{greedy_al} and \ref{greedy_al_nig} with $\beta=0,1$, naive smoothness detector, and BMSD by assuming the GMRF filter, with noise variance $\sigma_\wvec^2=0.2$, and $M=1,000$.}}
    \label{fig:ROCpower}
\end{figure}
 
\begin{figure}[hbt]
    \centering
    \includegraphics[width=7.75cm]{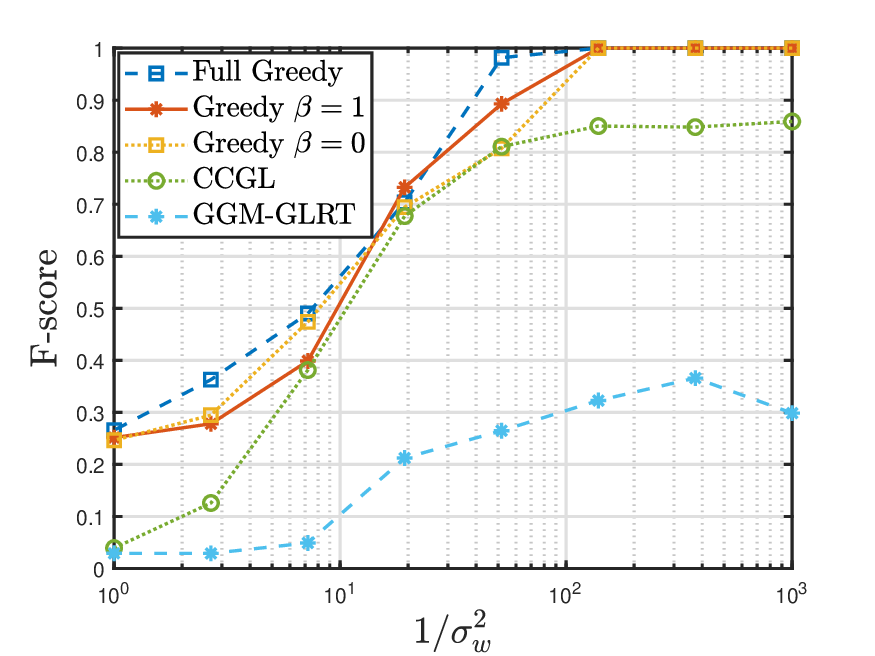}
    \caption{{\color{black}Identifying outages in power system: The F-score measure for power system dataset versus SNR, $1/ \sigma_{\wvec}^2$, for the following methods: greedy approaches from Algorithm \ref{greedy_al} and Algorithm \ref{greedy_al_nig} for $\beta=0,1$, which all assume GMRF filter with $M=10,000$.}}
    \label{fig:fscorepower}
\end{figure}

{\color{black} Figure \ref{fig:fscorepower}  presents the F-score measure  of the different methods versus $\frac{1}{\sigma^2_{\wvec}}$ for  $M=10,000$ time samples.  The CGL method  does not converge on this dataset, and therefore is not presented in this figure.
The simulation results  show that the proposed methods can be used for this practical scenario, while the GGM-GLRT has poor performance in this case.
In addition, the CCGL method performs well only for  a large enough number of time samples (for example, for $M=1,000$, the CCGL method does not converge).
}

\section{Conclusion}\label{Conclusion}
In this paper we investigate the problem of identifying edge disconnections in  networks based on a graph filter output model, where the initial topology of the graph is known. We show that the LRT for detecting   the status (connected/disconnected) of a specific set of edges can be interpreted as a local smoothness detector for the noiseless GMRF filter with a Laplacian precision matrix.
For the  general  edge disconnections identification problem, we show that the  ML decision rule   consists of parallel LRTs  of the binary hypothesis testing problems of each edge set. 
The interpretation of the LRT and the ML decision rule in the graph spectral domain is demonstrated.  
For nested subsets of edge disconnections and under mild conditions, the ML decision rule
imposes a larger penalty on models with a larger number of edge disconnections, which can be interpreted as a way to avoid overfitting.
In addition, it is shown that the sufficient statistics of the LRT and the ML decision rule are the graph energy
levels under each candidate graph. Thus, there is no need for an accurate estimation of the sample covariance matrix.
Since the ML  decision rule requires an exhaustive search and is impractical for large networks, we propose two greedy  algorithms for its low-complexity implementation. The greedy  solutions are iterative methods that are based on gaining a series of single-edge disconnection updates and on {\em{local}} properties of the smooth graph filters, by taking into account only the neighbors of the suspicious edges. 

Our simulations demonstrate that  the proposed methods outperform existing methods on the tested scenarios in terms of detection and identification performance, as well as computational complexity.  Thus, the proposed greedy methods  provide a good trade-off between performance and complexity and are practical for large-scale networks.
{\color{black} In addition, the simulations show the robustness of the  proposed neighboring strategy, which is based on the local property of the GMRF filter, for  other smooth graph filter models.
We also demonstrate the use of the proposed methods for detecting power system
line outages by using PMU phasor angle measurements.}
Future research directions include extension of the proposed methods for the case of disconnection identification with unknown graph filters, identification of edge disconnections in dynamic models, and detection of new connections in the networks. {\color{black}
A  deeper analysis of  the proposed framework  should be conducted, including determining the required number of measurements for sufficient performance and  analyzing the robustness of the results to the underlying assumptions.
}

\appendices
	\renewcommand{\thesectiondis}[2]{\Alph{section}:}
\section{Proof of Proposition \ref{eigenvalue}}\label{B}
We begin this appendix by proving that under the  conditions of  Proposition \ref{eigenvalue}, the eigenvalues of the considered Laplacian matrices  satisfy
    \be \label{eigenvalues pro}
    \lambda^{(\Lmat^{(k_1)})}_n\leq \lambda^{(\Lmat^{(k_2)})}_n,~~~ n=1,\ldots,N,\numberthis
    \ee
    where $\lambda_n^{(\Lmat^{(k_1)})}$ and $\lambda_n^{(\Lmat^{(k_2)})}$ are the $n$th eigenvalues of  $\Lmat^{(k_1)}$ and  $\Lmat^{(k_2)}$, respectively.
    By substituting the two Laplacian matrices in \eqref{multierror}-\eqref{multierror1}, one obtains
  \begin{align*}\label{multierror_app}
    \Lmat^{(k_t)}=\Lmat^{(0)}-\sum_{(i,j)\in {\mathcal{C}}^{(k_t)}  }\Emat^{(i,j)},\numberthis{}~t=1,2.
    \end{align*}
    Under Condition C.2 of Proposition \ref{eigenvalue},
 $\mathcal{C}^{(k_2)}$ is a proper subset of $\mathcal{C}^{(k_1)}$. 
 Thus,  \eqref{multierror_app} implies that
\begin{align*}
    \Lmat^{(k_2)} &= \Lmat^{(k_1)}+\sum_{(i,j)\in \mathcal{C}^{(k_1)}\setminus\mathcal{C}^{(k_2)}}\Emat^{(i,j)}.\numberthis
    \end{align*}
Since $\Lmat^{(k_1)}$, $\Lmat^{(k_2)}$, and $\sum_{(i,j)\in \mathcal{C}^{(k_1)}\setminus\mathcal{C}^{(k_2)}}\Emat^{(i,j)}$ are  Hermitian matrices, by using the Weyl's inequality  (see, e.g. Chapter 4.3 in \cite{Matrix}), one obtains
\begin{align*}\label{weyl}
    \lambda_n^{(\Lmat^{(k_1)})}+\tilde{\lambda}_1^E\leq\lambda_n^{(\Lmat^{(k_2)})},\quad 1\leq n\leq N,\numberthis{}
\end{align*}
where $\tilde{\lambda}_1^E$ is the smallest eigenvalue of $\sum_{(i,j)\in \mathcal{C}^{(k_1)}\setminus\mathcal{C}^{(k_2)}}\Emat^{(i,j)}$, which equals zero, since this matrix 
is a singular PSD matrix. Therefore,  by substituting $\tilde{\lambda}_1^{E}=0$ in (\ref{weyl}), we obtain  (\ref{eigenvalues pro}). It should be noted that a  (\ref{eigenvalues pro}) was proved  in \cite{Guo_2018} for normalized Laplacian matrices. 

Next, under Condition C.1 of Proposition \ref{eigenvalue}, the matrices $\bsigma(\Lmat^{(k_1)})$ and $\bsigma(\Lmat^{(k_2)})$ from (\ref{Sigma}) are non-singular matrices. Thus,  (\ref{rho p})  can be rewritten as  
\begin{align*}\label{MML 1n}
\rho(\Lmat^{(k)})&=\log\bigg(\prod_{n=1}^N\frac{\sigma_{\xvec}^2h^2(\lambda_n^{(\Lmat^{(k)})})+\sigma_{\wvec}^2}{\sigma_{\xvec}^2h^2(\lambda_n^{(\Lmat^{(0)})})+\sigma_{\wvec}^2}\bigg)\\
&=\sum_{n=1}^N\log\bigg(\frac{\sigma_{\xvec}^2h^2(\lambda_n^{(\Lmat^{(k)})})+\sigma_{\wvec}^2}{\sigma_{\xvec}^2h^2(\lambda_n^{(\Lmat^{(0)})})+\sigma_{\wvec}^2}\bigg).\numberthis
\end{align*}
Since  $\Lmat^{(k_1)}$ and $\Lmat^{(k_2)}$ represent connected graphs, the eigenvalues of these Laplacian matrices are positive, except for $\lambda_1^{(\Lmat^{(k_2)})}=\lambda_1^{(\Lmat^{(k_2)})}=0$. 
Thus, since the graph filter $h(\lambda)$ is a monotonic decreasing function for $\lambda$ for any $\lambda>0$ (Condition C.3 of Proposition \ref{eigenvalue}),  the order relation in (\ref{eigenvalues pro}) implies  that 
\begin{align*}\label{log_wn}
    \sigma_{\xvec}^2h^2(\lambda_n^{(\Lmat^{(k_1)})})\geq\sigma_{\xvec}^2h^2(\lambda_n^{(\Lmat^{(k_2)})}),\quad 1\leq n\leq N,\numberthis{}
\end{align*}
where we use the fact that $\sigma_{\xvec}^2>0$. By substituting  $k=k_1,k=k_2$ in (\ref{MML 1n}),
the inequality in \eqref{log_wn} implies that
\begin{align*}\label{p1}
    \rho(\Lmat^{k_2})
    &=\sum_{n=1}^N\log\bigg(\frac{\sigma_{\xvec}^2h^2(\lambda_n^{(\Lmat^{(k_2)})})+\sigma_{\wvec}^2}{\sigma_{\xvec}^2h^2(\lambda_n^{(\Lmat^{(0)})})+\sigma_{\wvec}^2}\bigg)\\
    &\leq\sum_{n=1}^N\log\bigg(\frac{\sigma_{\xvec}^2h^2(\lambda_n^{(\Lmat^{(k_1)})})+\sigma_{\wvec}^2}{\sigma_{\xvec}^2h^2(\lambda_n^{(\Lmat^{(0)})})+\sigma_{\wvec}^2}\bigg)=\rho(\Lmat^{k_1}),\numberthis{}
\end{align*}
which implies  (\ref{penalty pro}) and completes the proof of Proposition \ref{eigenvalue}.
\section{Development of $\rho(\Lmat^{(k)})$ for the noiseless GMRF filter}\label{C}
In this appendix, we prove that $\rho(\Lmat^{(k)})$ from (\ref{rho p}) for the noiseless GMRF filter in (\ref{GMRF filter})  is only a function of the
second-order statistics
 of the vertices in $\mathcal{S}^{(k)}$. By substituting (\ref{GMRF filter})  and $\sigma_{\wvec}^2=0$ in  (\ref{rho p}), we obtain
\be\label{rho GMRF}
    \rho(\Lmat^{(k)})=\log\bigg(\frac{|\sigma_{\xvec}^2\big(\Lmat^{(k)}\big)^\dagger|_+}{|\sigma_{\xvec}^2\big(\Lmat^{(0)}\big)^\dagger|_+}\bigg)
    =\log\bigg(\frac{|\Lmat^{(0)}|_+}{|\Lmat^{(k)}|_+}\bigg),
\ee
where the last equality stems from using the definition of the pseudo-determinant of a pseudo-inverse matrix and
\eqref{multierror}.
According to  Theorem 4 in \cite{generalized}, the Laplacian matrix of a connected graph  satisfies
\begin{align*}\label{ev}
|\Lmat^{(k)}|_+
=\prod_{n=2}^N\lambda_n^{(\Lmat^{(k)})}=
|\check{\Lmat}^{(k)}| \numberthis,
\end{align*}
where
\begin{align*}
\label{def_L_tilde}
    \check{\Lmat}^{(k)}\triangleq\Lmat^{(k)}+\frac{1}{N}\onevec\onevec^T\numberthis
\end{align*}
and $\lambda_2^{(\Lmat^{(k)})},\ldots,\lambda_N^{(\Lmat^{(k)})}$ are the eigenvalues of $\Lmat^{(k)}$. By substituting \eqref{ev} in \eqref{rho GMRF}, we obtain that in this case
\be\label{rho-GMRF2}
    \rho(\Lmat^{(k)})
    =\log\bigg(\frac{|\check{\Lmat}^{(0)}| }{|\check{\Lmat}^{(k)}| }\bigg).
\ee
In addition, 
according to  Theorem 5 in \cite{generalized}, the inverse of the matrix in  \eqref{def_L_tilde},
  satisfies
   \beqna
   \label{sig_L}
    (\check{\Lmat}^{(k)})^{-1}=(\Lmat^{(k)})^\dagger+\frac{1}{N}\onevec\onevec^T
    =\frac{1}{\sigma_{\xvec}^2}\bsigma(\Lmat^{(k)})+\frac{1}{N}\onevec\onevec^T,
\eeqna
where the last equality is obtained by substituting \eqref{GMRF_cov}, which holds under the GMRF model.

Without loss of generality, we assume that the non-zero elements in the matrix  $\Emat^{(k)}$ appear in the upper square block of $\Emat^{(k)}$, i.e.
the Laplacian matrix after the disconnections in $\mathcal{S}^{(k)}$ can be written as 
\begin{align*}\label{li4}
   \Lmat^{(k)}=\begin{bmatrix} \Lmat^{(0)}_{\mathcal{S}^{(k)}}-\Emat^{(k)}_{\mathcal{S}^{(k)}}&  \Lmat^{(0)}_{\mathcal{S}^{(k)},\Bar{\mathcal{S}}^{(k)}}&\\
   \Lmat^{(0)}_{\Bar{\mathcal{S}}^{(k)},\mathcal{S}^{(k)}}&\Lmat^{(0)}_{\Bar{\mathcal{S}}^{(k)}}&
   \end{bmatrix},\numberthis
\end{align*}
where $\Bar{\mathcal{S}}^{(k)}\triangleq\mathcal{V}\setminus \mathcal{S}^{(k)}$.
Similarly, according to \eqref{def_L_tilde}, we can write
\begin{align*}\label{li5}
   \check{\Lmat}^{(k)}=\begin{bmatrix} \check{\Lmat}^{(0)}_{\mathcal{S}^{(k)}}-\Emat^{(k)}_{\mathcal{S}^{(k)}}&  \check{\Lmat}^{(0)}_{\mathcal{S}^{(k)},\Bar{\mathcal{S}}^{(k)}}&\\
   \check{\Lmat}^{(0)}_{\Bar{\mathcal{S}}^{(k)},\mathcal{S}^{(k)}}&\check{\Lmat}^{(0)}_{\Bar{\mathcal{S}}^{(k)}}&
   \end{bmatrix},\numberthis
\end{align*}
where
$
    \check{\Lmat}^{(0)}\triangleq\Lmat^{(0)}+\frac{1}{N}\onevec\onevec^T$.
By applying  the block matrix determinant rule \cite{inverse}
on the matrix in \eqref{li5}, one obtains
\beqna\label{blockdet}
    | \check{\Lmat}^{(k)}|=
  |\check{\Lmat}^{(0)}_{\Bar{\mathcal{S}}^{(k)}}|\hspace{5.5cm}
    \nonumber\\
    \times|\check{\Lmat}^{(0)}_{\mathcal{S}^{(k)}}-\Emat^{(k)}_{\mathcal{S}^{(k)}}-\check{\Lmat}^{(0)}_{\mathcal{S}^{(k)},\Bar{\mathcal{S}}^{(k)}}(\check{\Lmat}^{(0)}_{\Bar{\mathcal{S}}^{(k)}})^{-1}\check{\Lmat}^{(0)}_{\Bar{\mathcal{S}}^{(k)},\mathcal{S}^{(k)}}|.
\eeqna
Then, by using the partitioned matrix inversion lemma (see, e.g. Eq. (8) in \cite{inverse}) on
\eqref{li5}, we obtain that the upper block of the inverse matrix of $\check{\Lmat}^{(k)}$ satisfies
\beqna\label{part_inverse_l}
 [(\check{\Lmat}^{(k)})^{-1}]_{\mathcal{S}^{(k)}}=\hspace{5.5cm}\nonumber\\
 \big(\check{\Lmat}^{(0)}_{\mathcal{S}^{(k)}}-\Emat^{(k)}_{\mathcal{S}^{(k)}}-\check{\Lmat}^{(0)}_{\mathcal{S}^{(k)},\Bar{\mathcal{S}}^{(k)}}(\check{\Lmat}^{(0)}_{\Bar{\mathcal{S}}^{(k)}})^{-1}\check{\Lmat}^{(0)}_{\Bar{\mathcal{S}}^{(k)},\mathcal{S}^{(k)}}\big)^{-1}.
\eeqna
By substituting \eqref{part_inverse_l}  in \eqref{blockdet}, one obtains
\beqna\label{blockdet22}
    | \check{\Lmat}^{(k)}|=
  |\check{\Lmat}^{(0)}_{\Bar{\mathcal{S}}^{(k)}}|
  \left|\left([(\check{\Lmat}^{(k)})^{-1}]_{\mathcal{S}^{(k)}}\right)^{-1}\right|\hspace{2.1cm}\nonumber\\
    =|\check{\Lmat}^{(0)}_{\Bar{\mathcal{S}}^{(k)}}|
  \left|
  \left(\frac{1}{\sigma_{\xvec}^2}[\bsigma(\Lmat^{(k)})]_{\mathcal{S}^{(k)}}+\frac{1}{N}\onevec\onevec^T\right)^{-1}\right|,
\eeqna
where the last equality is obtained by using \eqref{sig_L}.
In a similar manner to the development of \eqref{blockdet22}, it can be shown that  the determinant of $\check{\Lmat}^{(0)}$ satisfies
\beqna\label{blockdet3}
    | \check{\Lmat}^{(0)}|=
 |\check{\Lmat}^{(0)}_{\Bar{\mathcal{S}}^{(k)}}|
  \left|
  \left(\frac{1}{\sigma_{\xvec}^2}[\bsigma(\Lmat^{(0)})]_{\mathcal{S}^{(k)}}+\frac{1}{N}\onevec\onevec^T\right)^{-1}\right|.
\eeqna
By substituting (\ref{blockdet22}) and (\ref{blockdet3})  in \eqref{rho-GMRF2}, we obtain
\begin{align*}\label{smooth measure1}
   \rho(\Lmat^{(k)})&=\log\bigg(\frac{|\check{\Lmat}^{(0)}_{\Bar{\mathcal{S}}^{(k)}}|
  |
  (\frac{1}{\sigma_{\xvec}^2}[\bsigma(\Lmat^{(0)})]_{\mathcal{S}^{(k)}}+\frac{1}{N}\onevec\onevec^T)^{-1}|}{|\check{\Lmat}^{(0)}_{\Bar{\mathcal{S}}^{(k)}}|
  |
  (\frac{1}{\sigma_{\xvec}^2}[\bsigma(\Lmat^{(k)})]_{\mathcal{S}^{(k)}}+\frac{1}{N}\onevec\onevec^T)^{-1}|}\bigg)\\
   &=\log\bigg(\frac{| \frac{1}{\sigma_{\xvec}^2}[\bsigma(\Lmat^{(k)})]_{\mathcal{S}^{(k)}}+\frac{1}{N}\onevec\onevec^T|}{| \frac{1}{\sigma_{\xvec}^2}[\bsigma(\Lmat^{(0)})]_{\mathcal{S}^{(k)}}+\frac{1}{N}\onevec\onevec^T|}\bigg)\numberthis{}.
\end{align*}
That is, we obtained that $\rho(\Lmat^{(k)})$ for the noiseless GMRF filter is only a function of the second-order statistics of the vertices in $\mathcal{S}^{(k)}$, as required.
\bibliographystyle{IEEEtran}

\end{document}